\documentclass[superscriptaddress,showpacs,preprintnumbers,showkeys,amsmath,amssymb,floatfix]{revtex4}
\usepackage{color}
\usepackage{graphicx}
\usepackage{dcolumn}
\usepackage{bm}
\usepackage[mathscr]{eucal}
\usepackage[nooneline]{subfigure}

\newcommand{\etal}{\textit{et al}}
\newcommand{\ud}{\mathrm{d}}

\newcommand{\bra}[1]{\langle#1\vert}
\newcommand{\ket}[1]{\vert#1\rangle}
\newcommand{\op}[2]{\ket{#1}\bra{#2}}
\newcommand{\inner}[2]{\bra{#1} #2\rangle}
\newcommand{\braket}[1]{\langle #1 \rangle}

\newcommand{\eqn}[1]{Eq.~(\ref{#1})}
\newcommand{\eqns}[1]{Eqs.~(\ref{#1})}
\newcommand{\ignore}[1]{}

\newcommand{\commute}[2]{[#1,#2]}

\DeclareMathOperator{\Tr}{Tr}

\newcommand{\fig}[1]{Fig.~\ref{#1}}

\newcommand{\jump}{{\mathcal J}}
\newcommand{\A}{{\mathcal A}}

\newcommand{\hc}{\mathsf{H.c.}}
\newcommand{\n}{\hat n}

\newcommand{\heff}{\mathcal{H}}
\newcommand{\ueff}{\mathcal{U}}
\newcommand{\Boltz}{k_B}
\renewcommand{\vec}[1]{\mathbf{#1}}
\usepackage{hyperref}

\newcommand{\singlet}{S}
\newcommand{\tripup}{T_{\uparrow\uparrow}}
\newcommand{\tripdown}{T_{\downarrow\downarrow}}
\newcommand{\tripud}{T_{\uparrow\downarrow}}
\newcommand{\Dplus}{D_+}
\newcommand{\Dminus}{D_-}

\newcommand{\DoubleOccupationEnergy}{U}
\newcommand{\isle}{s}
\newcommand{\Coulomb}{V}
\newcommand{\hhub}{H_\mathrm{Hub}}
\newcommand{\hmeas}{H_\mathrm{meas}}
\newcommand{\hsys}{H_\mathrm{sys}}
\newcommand{\hrot}{H_\mathrm{0}}
\newcommand{\hi}{H_\mathrm{I}}
\newcommand{\htot}{H_\mathrm{Tot}}
\newcommand{\htun}{H_\mathrm{tun}}
\newcommand{\hsp}{H_\mathrm{Hub}}
\newcommand{\spenergydiff}{\Delta}
\newcommand{\hleads}{H_\mathrm{leads}}
\newcommand{\hisle}{H_\mathrm{island}}
\newcommand{\DWS}{DWS}
\newcommand{\rhoi}{R_\mathrm{I}}
\newcommand{\rhos}{R}
\newcommand{\rhodw}{\rho}
\newcommand{\dummy}{t'}
\newcommand{\lind}{\mathcal{D}}
\newcommand{\coula}{a}
\newcommand{\coulb}{\delta}
\newcommand{\coulc}{\epsilon}
\newcommand{\diracdelta}{\delta}

\newcommand{\micro}{$\mu$}
\newcommand{\ns}{{n_s}}
\newcommand{\coulcoeff}{q}

\newcommand{\InverseBohrRadius}{\mu}
\newcommand{\DonorSep}{d}
\newcommand{\leadS}{l}
\newcommand{\leadD}{r}
\newcommand{\tmeas}{t_\mathrm{meas}}
\newcommand{\tmix}{t_\mathrm{mix}}
\newcommand{\prob}{\mathcal{P}}

\newcommand{\gres}{G_{\mathrm{res}}}
\newcommand{\gcot}{G_{\mathrm{cot}}}

\begin{document}


\title{Parity measurement of one- and two-electron double well systems.}

\author{T.M. Stace}
\affiliation{Cavendish Laboratory, University of Cambridge,  Madingley Road, Cambridge CB3 0HE, UK}
\affiliation{DAMTP, University of Cambridge, Wilberforce Rd CB30WA, UK}
\email{tms29@cam.ac.uk}
\author{S. D. Barrett}
\affiliation{Hewlett Packard Laboratories,  Filton Road, Stoke Gifford Bristol, BS34 8QZ}
\author{H-S. Goan}
\affiliation{Centre for Quantum Computer Technology,
University of New South Wales, Sydney, NSW 2052, Australia}
\author{G.J. Milburn}
\affiliation{Centre for Quantum Computer Technology,
University of Queensland, St Lucia, QLD 4072, Australia}

\date{\today}



\begin{abstract}
We outline a scheme to accomplish measurements of a solid state double well system (\DWS) with both one and two electrons in non-localised bases.  We show that, for a single particle, measuring the local charge distribution at the midpoint of a \DWS\ using an SET as a sensitive electrometer amounts to performing a projective measurement in the parity (symmetric/antisymmetric) eigenbasis.  For two-electrons in a \DWS, a similar configuration of SET  results in close-to-projective measurement in the singlet/triplet basis.  We analyse the sensitivity of the scheme to asymmetry in the SET position for some experimentally relevant parameter, and show that it is realisable in experiment.

\end{abstract}
\maketitle
\section{Introduction}

In this paper, we present a scheme for performing measurements of one- and two-electron double well systems (\DWS s) in non-local bases.  The principle idea behind both schemes is to detect the small charge differences at the midpoint of the \DWS\ for different electronic states using a single electron transistor (SET) as a sensitive electrometer.  SET's have been demonstrated to be highly sensitive electrometers, with sensitivities of a few $\mu e/\sqrt{\mathrm{Hz}}$ \cite{aas01,mak01,bue03}.  We therefore expect that they could detect fluctuations of around 1\% of an electron charge in around 10-100 ns.  

For the one electron case, the measurement is in the parity eigenbasis.  This is interesting since it results in projection onto non-local states.  It has been shown theoretically \cite{sta04} and experimentally \cite{hay03} that decoherence is slower for an evenly biased \DWS, so this kind of measurement may be more robust against decoherence.   

For a two electron system we show that this measurement approximately projects the \DWS\ onto the singlet (even) and triplet (odd) subspaces.  It is therefore a method for performing spin sensitive detection using electrometers, which is important for readout of certain quantum information processing schemes \cite{div00}.

The paper begins in section \ref{sec:system} with a short, generic discussion of the microscopic model of a few electron system interacting with an idealised SET, itself in contact with a continuum of lead modes.  Following this, section  \ref{sec:SP} deals with the single electron case, and section \ref{sec:TP} deals with the two electron case.  Within each of these two sections we develop the measurement Hamiltonian from microscopic considerations, from which we derive measurement and mixing times. 
 After showing in each case that the measurements work in principle,  we estimate the effects of a significant problem in the fabrication of this device, namely the precision with which the SET island must be placed in the midplane of the \DWS.  Section \ref{sec:estimationofparameters} concludes the paper with a discussion about experimental implementation of the scheme.

\section{System}\label{sec:system}

\begin{figure}
\includegraphics[width=4cm]{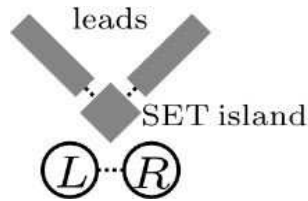}
\caption{\label{fig:DoubleWellSystem} Schematic of physical system under consideration}
\end{figure}

We consider the device pictured in \fig{fig:DoubleWellSystem}, consisting of a SET placed in the mid-plane of the \DWS, in order to be sensitive to the charge at the midpoint.  This charge differs between symmetric and antisymmetric spatial wavefunctions, and we analyse a scheme to measure this difference in order to effect projective measurements onto the parity eigenspaces.

For the purposes of this paper, we assume the SET island has only a single accessible energy level, which is reasonable if the island is small and the difference between Fermi energies in the leads is less than the charging energy of the island.  We model the \DWS\ with a Hubbard Hamiltonian, with only a single spatial wavefunction per well, $\ket{L}$ and $\ket{R}$ for the left and right wells.  We assume that the Hilbert space for the system is therefore two-dimensional, which is reasonable if higher excited state are inaccessible due to the strong confinement of the quantum dot potentials.  Therefore, each well may be populated by at most two electrons, in different spin configurations. 

The Hamiltonian for a system of interacting electrons is given by
\begin{equation}
\htot=\sum_{i j,\sigma} H_{i j}c^\dagger_{i\sigma} c_{j\sigma}+
\frac{1}{2}\sum_{i j l m,\sigma \sigma'}
\Coulomb_{ijmn}c^\dagger_{j\sigma} c^\dagger_{n\sigma'}c_{m\sigma'}c_{i\sigma},
\label{eqn:Hamiltonian1}
\end{equation}
where $c_{i,\sigma}$ is a (fermionic) annihilation operator for an electron in spatial mode $i$ and spin $\sigma\in\{\uparrow,\downarrow\}$ \cite{mahan} and
\begin{eqnarray}
H_{ij}&=&\int \ud^3\vec{r}\phi_i^*(\vec{r})\left(-\hbar^2\nabla^2/2m +U(\vec{r})\right)\phi_j(\vec{r})\\
\Coulomb_{ijmn}&=&\int \ud^3\vec{r}_1\int \ud^3\vec{r}_2 \phi_j^*(\vec{r}_1)\phi_i(\vec{r}_1)V(|\vec{r}_1-\vec{r}_2|) \phi_n^*(\vec{r}_2)\phi_m(\vec{r}_2),\label{eqn:coulint}
\end{eqnarray}
where $\phi_i(\vec{r})=\bra{\vec{r}}c^\dagger_{i\sigma}\ket{}$ is the spatial wavefunction for mode $i$ (assuming both spin states have the same spatial wavefunction), $\ket{}$ is the `vacuum' state, with no quasi-particle excitations, $m$ is the effective mass, $U(r)$ is the electrostatic confining potential and $V(r)=\frac{e^2}{4\pi \varepsilon_0}\frac{1}{r}=\frac{\coulcoeff}{r}$ is the Coulomb potential.

We choose a basis set such that $H_{i j}$ is diagonal, which we truncate to the lowest two eigenstates for the \DWS, and a single state, $\isle$, on the SET.  In the absence of an external bias between the wells, this corresponds to taking $i,j\in \{+,-,s\}$, where $\pm$ are the symmetric and antisymmetric superpositions of the localised, single-particle states, referred to here as the parity eigenbasis.  That is 
$\ket{\pm}=c^\dagger_\pm\ket{}=\frac{\ket{L}\pm\ket{R}}{\sqrt{2(1\pm\inner{R}{L})}}.$
The first term of \eqn{eqn:Hamiltonian1} becomes 
$\sum_{\sigma}\frac{\spenergydiff}{2}(\n_{+\sigma}-\n_{-\sigma})+\omega_0\n_\isle$, where $\n_{i\sigma}=c^\dagger_{i\sigma} c_{i\sigma}$ and $\spenergydiff=H_{++} - H_{--}$ is the tunnelling rate of a localised electron, which we estimate in section \ref{sec:estimationofparameters}.

Expanding \eqn{eqn:Hamiltonian1} gives \begin{equation}\htot=\hisle+\hleads+\htun+\hhub+\hat\Coulomb,\end{equation}
 where $\hhub$ is the Hubbard Hamiltonian for the \DWS, $\hat\Coulomb$ is the Hamiltonian for the interaction between the \DWS\ and the SET and
\begin{eqnarray}
\hisle&=&\omega_0 \n_\isle,\label{eqn:hisle}\\
\hleads&=&\sum_k \omega_k(\n_{l k}+\n_{r k}),\label{eqn:hleads}\\
\htun&=&\sum_k T_{l k} c^\dagger_{l k} c_\isle+T_{r k} c^\dagger_{r k} c_\isle+\hc,\label{eqn:htun}
\end{eqnarray}
Here $\omega_0$ is the island energy level, in the absence of interactions with the double well potential, $\omega_k$ are the energies of densely spaced lead modes, $l$ and $r$ denote the left and right leads respectively, $T_{l(r) k}$ are the corresponding tunnelling rates between mode $k$ in lead $l(r)$ and the SET island.

To compute $\hat \Coulomb$  we assume that the wavefunction for electrons on the SET island vanishes in the region where the wavefunction for the electron on the \DWS\ has support, and vice versa, so that for instance $\phi^*_+(\vec{r})\phi_s(\vec{r})=0$.  This assumption is a good one for systems such as the Kane proposal \cite{kan98}, or Na$^+$ in Si \cite{NaSi}, where the tunnelling rates between SET island and the \DWS\ are negligible.  

The consequence of this assumption is that if any index in $V_{ijmn}$ is $\isle$, 
 then $V_{ijmn}$ is zero unless $i=j=\isle$ or $m=n=\isle$, where $\isle$ labels an electron on the SET island.  Therefore the only Coulomb terms that contribute to the interaction between the SET island and the \DWS\ are given by
\begin{equation}
\hat \Coulomb=\sum_{i,j\in\{+,-\}}\Coulomb_{\isle \isle i j} \n_\isle c^\dagger_{i\sigma} c_{j\sigma}=\n_\isle\otimes\hmeas\label{eqn:paritycoulomb},
\end{equation}
where we have ignored the spin degree of freedom on the SET island.  There are four distinct terms of this form that need to be included, for $i,j$ taking the four possible assignments of $+$ or $-$.  If the physical arrangement of double well and island as shown in \fig{fig:DoubleWellSystem} is symmetric about a line bisecting the double well potential, then $\coulc\equiv \Coulomb_{\isle \isle +-}=\Coulomb_{\isle \isle-+}=0$, as discussed in appendix \ref{app:approximations}.  Asymmetry results in non-zero $\coulc$, and for the development of this section, we assume that it is symmetric, so that $\coulc=0$.  
Thus, to describe a \DWS\ interacting with an SET we need to specify $\hhub$ and $\hmeas$.

\section{Parity measurement for singly occupied \DWS}\label{sec:SP}

The system we consider in this section consists of a single electron shared between two wells, so we ignore spin indices.  We now establish the feasibility of performing a measurement in the parity eigenbasis of a single electron shared between the two wells.

\subsection{Derivation of measurement Hamiltonian}\label{sec:paritysingleham}

For a single electron on an unbiased double well, the interaction terms of \eqn{eqn:Hamiltonian1} vanishes, so the Hubbard Hamiltonian for the single particle system is given by
\begin{equation}
\hsp=\spenergydiff
 \n_+.\label{eqn:SingleParticleHam}
\end{equation}
 We have used the fact that, for a single-particle, two-level system $\n_++\n_-=I$ is the identity, and have also discarded terms proportional to $I$.  In an alternate notation, we identify $\n_+$ with the $\sigma_x=\op{L}{R}+\op{R}{L}$ operator.

For the symmetric case,
\begin{equation}
\hat \Coulomb=\n_\isle\otimes(\Coulomb_{\isle \isle + +}  \n_+ +  \Coulomb_{\isle \isle - -}  \n_-) =\coulb\n_\isle\n_+,\label{eqn:SingleParticleMeas}
\end{equation}
where $\coulb=\Coulomb_{\isle \isle + +}-\Coulomb_{\isle \isle - -}$ and we have again used the single particle, two-level system identity $\n_++\n_-=I$
.  It is evident from \eqns{eqn:SingleParticleHam} and (\ref{eqn:SingleParticleMeas}) that the measurement Hamiltonian and the system Hamiltonian have the same energy eigenstates.  Therefore, the measurement process will non-destructively project onto energy eigenstates of the \DWS, which are the delocalised symmetric and antisymmetric wavefunctions.  That is, it is a QND (quantum non-demolition) measurement, which simplifies the analysis greatly.

\subsection{Master equation for symmetric system}

As discussed in appendix \ref{app:mesp}, the master equation for the \DWS\ and SET island is given by
\begin{equation}
\dot \rhos(t)={}-i \commute{\hsp+\hisle}{\rhos(t)}
+(\gamma_l+\gamma_r )\lind[\n_-c_\isle^\dagger]\rhos(t)
+\gamma_r'\lind[\n_+c_\isle]\rhos(t)
+\gamma_l'\lind[\n_+c_\isle^\dagger]\rhos(t),\label{eqn:SPMaster1}
\end{equation}
where $\rhos$ is the density matrix for the  \DWS\ and SET island and $\lind[A]B\equiv \jump[A]B-\A[A]B\equiv ABA^\dagger-\frac{1}{2}(A^\dagger A B+B A^\dagger A)$. 
We assume that the reduced state of the SET island is diagonal in the number representation, so
\begin{equation}
\rhos\rightarrow\rhodw_0\otimes\ket{0}_s\bra{0}+\rhodw_1\otimes\ket{1}_s\bra{1},\label{eqn:R}
\end{equation}
The reduced density matrix of the DWS is given by $\rho(t)=\Tr_\isle\{R(t)\}=\rho_0(t)+\rho_1(t)$.  In appendix \ref{app:mesp} we solve \eqn{eqn:SPMaster1} and the steady-state reduced density matrix of the \DWS\ is 
\begin{equation}
\Tr_\isle\{R(\infty)\}=\rhodw(\infty)=
(\rhodw_0^{++}(0)+\rhodw_1^{++}(0))\n_+ +(\rhodw_0^{--}(0)+\rhodw_1^{--}(0))\n_-,
\end{equation}
where $\rhodw^{pq}=\bra{p}\rhodw\ket{q}$.  This shows that the diagonal elements of the system density matrix are unchanged, whilst the off-diagonal elements have completely decayed, which corresponds to QND measurement.

\subsection{Measurement time}\label{sec:SPmeastime}

\begin{figure}
\includegraphics{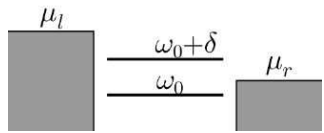}
\caption{SET island energies relative to lead Fermi level, depending on state of \DWS.}
\label{fig:SingleParticleEnergyLevel}
\end{figure}

Since the measurement for the single electron case is a QND measurement, we can straightforwardly calculate the expected currents and measurement times for the device.  If we choose $f_l(\omega_0)=f_l(\omega_0+\coulb)=1=f_r(\omega_0)$ and $f_r(\omega_0+\coulb)=0$, as shown in \fig{fig:SingleParticleEnergyLevel}, then the current through the SET is sensitive to the state of the \DWS. In this configuration, the SET island is  conducting when the \DWS\ is in a symmetric state, so a current $i_+$ flows, and non-conducting when in an antisymmetric state, so $i_-=0$.  Thus the measurement amounts to distinguishing the currents $i_+$ and $i_-=0$ through the SET.

For the symmetric configuration, the rates at which electrons hop on and off the SET island, are given by $\gamma_l$ and $\gamma_r$.  Thus the rate of transport of electrons through the SET island is $1/(\gamma_l^{-1}+\gamma_r^{-1})$. The current is therefore $i_+=e/(\gamma_l^{-1}+\gamma_r^{-1})$.  

The measurement time is then the time required to distinguish two currents, $i_+$ from $i_-=0$ in the presence of shot noise.  Since the transport of electrons through the SET is Poissonian, the probability of detecting zero electrons tunnelling in a time $\tau$ through the SET island, given that a mean current $i_+$ is flowing is given by $P_0(T)=e^{\tau/(\gamma_l^{-1}+\gamma_r^{-1})}$.  We therefore conclude that the measurement time is $\tmeas\approx\gamma_l^{-1}+\gamma_r^{-1}$, since the probability of not detecting a tunnelling event in  (a few multiples of) this time is small.  This agrees with the measurement determined from the decay rate of off diagonal elements of the system density matrix, found in appendix \ref{app:mesp}.

\subsection{Effects of asymmetry}\label{sec:parityasym}

\begin{figure}
\includegraphics[width=3.5cm]{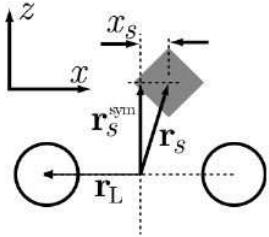}
\caption{Geometry of asymmetric system.}
\label{fig:AsymmetricSystem}
\end{figure}

We now turn to the important practical issue of how precisely the SET island needs to be placed with respect to the centre of the double well potential.  Here we assume there is some asymmetry, which may, for example, arise from fabrication, so that $\coulc=V_{ss+-}\neq0$.  We will estimate the magnitude of this quantity later, but first we will determine the effect of the extra terms in the Hamiltonian that arise.  Including this term in $\hat \Coulomb$ gives
\begin{equation}
\hat \Coulomb=\n_\isle\otimes\left(\coulb\n_++\coulc(c_+^\dagger c_- + c_-^\dagger c_+)\right).\label{eqn:asympot}
\end{equation}
This shows that the asymmetry rotates the measurement basis by an angle $\phi=\tan^{-1}(\coulc/\coulb)$ away from the parity eigenbasis, that is, the preferred basis for the measurement is $\{\cos(\phi)\ket{+}+\sin(\phi)\ket{-},-\sin(\phi)\ket{+}+\cos(\phi)\ket{-}\}$.  We therefore require that $\coulc\ll\coulb$ in order that the asymmetry have negligible effect.

In Appendix \ref{app:approximations} we estimate $\coulc$ and $\coulb$ to be 
\begin{equation}
\coulc\approx \coulcoeff\frac{2 x_\isle x_L}{|\vec{r_L}-\vec{r}_\isle^\mathrm{\,sym}|^3}\lesssim \coulcoeff\frac{2 x_\isle}{|\vec{r_L}-\vec{r}_\isle^\mathrm{\,sym}|^2}\textrm{ and }\coulb\approx \coulcoeff\inner{L}{R}\left(\frac{1}{|\vec{r}_\isle^\mathrm{\,sym}|}-\frac{1}{|\vec{r}_L-\vec{r}_\isle^\mathrm{\,sym}|}\right),
\end{equation}
where $\vec{r_L}$ is the `centre-of-mass' of the left well,  $\vec{r}_\isle=\{x_\isle,y_\isle,z_\isle\}$ is the position vector of the SET island and we choose the origin to be at the midpoint of the \DWS.  Note that $x_\isle=0$ for a symmetrically placed island, and is assumed to be small.

The condition that $\coulc\ll\coulb$ is therefore satisfied if $\frac{\coulc}{\coulb}\approx \frac{1}{\inner{L}{R}}\frac{2 x_\isle |\vec{r}_\isle^\mathrm{\,sym}|}{|\vec{r}_L-\vec{r}_\isle^\mathrm{\,sym}|^2}\ll1$.  This is a tight constraint, since it requires that the asymmetry, quantified by $\frac{2 x_\isle |\vec{r}_\isle^\mathrm{\,sym}|}{|\vec{r}_L-\vec{r}_\isle^\mathrm{\,sym}|^2}$, be much less than the overlap of the localised wavefunctions $\inner{L}{R}$.  With the help of a $J$-gate (as referred to in \citet{kan98}) $\inner{L}{R}$ may be made as high as 0.03 \cite{bur99}, and assuming a typical scale of device of $|\vec{r}_L-\vec{r}_\isle|\sim30$ nm, the elements of the SET and \DWS\ would likely need to be made with a precision of 1 nm or less, which seems difficult with current technology.

This issue may not be so significant for electrostatically defined dots, since the position of the SET island and \DWS\ may be changed by variation of surface gate potentials.  It is a serious problem for donor systems with SET's grown by metallic deposition, since the location of the donor atoms and SET island are fixed during fabrication.

\subsection{Mixing time}\label{sec:SPmixingtime}

Asymmetry in the placement of the SET island induces mixing in the state of the \DWS, so there is a mixing time associated with asymmetry.  The calculation of the mixing time is somewhat lengthy, but not difficult.  We derive an unconditional master equation for the density matrix of the \DWS\ and SET, $R$.  The solution to the master equation has exponentially decaying terms, with different time constants.  For the sake of simplicity, here we present the results of the calculation, and leave the details to appendix \ref{app:asymmesp}, which follows from the results of appendix \ref{app:noncomME}.  Taking $\gamma_i=\gamma$, the most rapidly decaying term gives the measurement time, $\tmeas={2}/{\gamma}$, which is unchanged from the symmetric case (to within $O(\coulc^2)$).  The slowest decaying term gives the measurement induced mixing time, $\tmix=\frac{(\coulb+\spenergydiff)^2}{\coulc^2}\frac{2}{3\gamma}$, so the condition for a good measurement is that $\tmix\gg\tmeas$, which occurs if $\coulb+\spenergydiff\gg\coulc$, in agreement with above.  When $\coulc=0$, the measurement is QND, $\tmix=\infty$.

When $\coulc\neq0$, the dynamics are divided into two regimes, $t\ll\tmix$ and $t\sim\tmix$.  In the energy eigenbasis, for short times, $t\ll\tmix$,  the diagonal elements of the density matrix are almost unaffected whilst the off-diagonal elements decay at a rate $\gamma/2$.  This corresponds to the process of projecting the \DWS\  onto its energy eigenstates.  
Over longer times, $t\sim\tmix$, the diagonal matrix elements decay to their steady-state values, given by \eqn{eqn:sssolsp}, corresponding to almost complete relaxation of the \DWS\ to its ground state $\ket{-}$.  

\subsection{Summary}

Performing a `non-local' measurement by projecting onto the parity eigenbasis of a \DWS\ is in principle possible.  An SET placed in the midplane of the \DWS\  is sensitive to the local charge at the midpoint of the system, which depends on the parity of the state.

Estimates of the sensitivity of such a device to asymmetry in the placement of the SET island suggest that it would need to be placed with a precision better than $\inner{L}{R}\approx3\%$ of the typical device dimensions.  For a donor based system, this is around 1 nm, which has been demonstrated recently for P$^+$ donors in Si \cite{sch03}.

The scheme may still be usefully applied in the case where $\coulc\lesssim\coulb$, where the preferred measurement basis is close to the parity eigenbasis.  Applying a bias across the \DWS, so that one well is at a higher potential than the other, would allow one to rotate the double well energy eigenbasis onto the measurement eigenbasis. 

The precise placement of the SET island in an electrostatically defined system may not present serious problems, since it may be moved about after fabrication by varying surface gate voltages.  In section \ref{sec:estimationofparameters}, we give estimates of expirementally accessible parameter values, and show that this proposal is experimentally viable.

\section{Singlet/Triplet measurement for doubly occupied \DWS}\label{sec:TP}

We consider here a \DWS\ populated with two electrons, shown schematically in \fig{fig:DoubleWellSystem}.  As in the single electron parity measurement scheme, the SET island  is placed in the midplane of the \DWS, so that it is sensitive to the charge distribution on the \DWS.  

The physical principle that we exploit is the fact that a pair of electrons in a triplet state are Pauli blocked from being simultaneously at the origin of the \DWS, so the probability amplitude to find two electrons in a triplet at the origin is zero.  In contrast, this restriction does not apply to a pair in a singlet state, so there is a non-zero probability amplitude to find two electrons in a singlet near the origin.  Thus there is a small variation in the local charge density at the origin between singlet and triplet states, which can in principle be measured to distinguish these subspaces.  This has some similarities to another singlet-triplet measurement scheme \cite{SeansPaper}.

Such systems may be used to implement quantum information processing tasks.  In certain instances it is important to distinguish whether the two electrons are in a singlet state or a triplet state, thereby providing information about their spin state, e.g. distinguishing a  state from the triplet states is necessary in the three-in-one encoding scheme developed by  \citet{div00}.

\subsection{Derivation of measurement Hamiltonian}

In this system, there is one singly occupied singlet state and three singly occupied triplet states, which are given by
\begin{eqnarray}
\ket{\singlet}&=&\frac{1}{\sqrt{2}}(c^\dagger_{+\uparrow} c^\dagger_{+\downarrow}-c^\dagger_{-\uparrow} c^\dagger_{-\downarrow})\ket{},\\
\ket{\tripup}&=&c^\dagger_{+\uparrow}c^\dagger_{-\uparrow}\ket{},\\
\ket{\tripdown}&=&c^\dagger_{+\downarrow}c^\dagger_{-\downarrow}\ket{},\\
\ket{\tripud}&=&\frac{1}{\sqrt{2}}(c^\dagger_{-\uparrow} c^\dagger_{+\downarrow}-c^\dagger_{+\uparrow} c^\dagger_{-\downarrow})\ket{}.
\end{eqnarray}  
The spatial wavefunction of the singlet state is clearly symmetric, whilst the spin wavefunction is antisymmetric.  The converse is true for the triplet states.  Written in the parity eigenbasis, it is clear that there are different charge densities between the wells depending on the state: the singlet state is a superposition of terms with non-zero amplitude to find either zero electrons (both in the $\ket{-}$ state, with zero charge density at the midpoint) or two electrons (both in the $\ket{+}$ state, with non-zero charge density at the midpoint) to exist between the wells, whilst the triplet states have an amplitude to find only a single electron (only one electron in the $\ket{+}$) state to be located at the midpoint.     
 There are also two doubly occupied states given by 
\begin{eqnarray}
\ket{\Dplus}&=&\frac{1}{\sqrt{2}}(c^\dagger_{+\uparrow} c^\dagger_{+\downarrow}+c^\dagger_{-\uparrow} c^\dagger_{-\downarrow})\ket{},\\
\ket{\Dminus}&=&\frac{1}{\sqrt{2}}(c^\dagger_{-\uparrow} c^\dagger_{+\downarrow}+c^\dagger_{+\uparrow} c^\dagger_{-\downarrow})\ket{}.
\end{eqnarray}

As discussed in section \ref{sec:system}, we need to specify the \DWS\ dynamics, given by $\hhub$, as well as the interaction between the \DWS\ and the SET, given by $\hmeas$. 
In the parity eigenbasis, $\hhub$ is given by 
\begin{eqnarray}
\hhub=&& \spenergydiff/2 (\n_{+\uparrow}+\n_{+\downarrow}-\n_{-\uparrow}-\n_{-\downarrow})\nonumber\\&+&{ \DoubleOccupationEnergy}/{2}\left((\n_{+\uparrow}+\n_{-\uparrow})(\n_{+\downarrow}+\n_{-\downarrow})+
(c^\dagger_{+\uparrow}c_{-\uparrow}+c^\dagger_{-\uparrow}c_{+\uparrow})(c^\dagger_{+\downarrow}c_{-\downarrow}+c^\dagger_{-\downarrow}c_{+\downarrow})\right),
\end{eqnarray}
where $\spenergydiff=H_{++} - H_{--}$ as defined earlier and 
$ \DoubleOccupationEnergy=V_{LLLL}$ is the double occupation Coulomb energy.  The triplet states are eigenstates of the two-site Hubbard Hamiltonian, so decouple from the other states.  With respect to the ordered sub-basis $\{\ket{\singlet},\ket{\Dplus},\ket{\Dminus}\}$ the matrix for the Hubbard Hamiltonian is 
\begin{equation}
\hhub=\begin{bmatrix}
     -\DoubleOccupationEnergy & \spenergydiff & 0  \\
     \spenergydiff & 0 & 0\\
     0 & 0 & 0
\end{bmatrix}
\end{equation}

We now turn to the interaction between the two electron system and the nearby SET island populated with at most one electron.  There are four distinct terms in \eqn{eqn:paritycoulomb} that need to be computed, for $i,j$ taking the four possible assignments of $+$ or $-$.  The triplet states are once again eigenstates of $\hmeas$, with eigenvalue $\Coulomb_{ss++}+\Coulomb_{ss--}$.  Again with respect to the ordered sub-basis $\{\ket{\singlet},\ket{\Dplus},\ket{\Dminus}\}$, the matrix representation of $\hmeas$ is 
\begin{equation}
\hmeas=
\begin{bmatrix}
     \coula & \coulb & 0  \\
     \coulb & a & 2\coulc\\
     0 & 2\coulc & \coula
\end{bmatrix},\label{eqn:hmeas}
\end{equation}
where $\coula=\Coulomb_{ss++}+\Coulomb_{ss--}$, $\coulb=\Coulomb_{ss++}-\Coulomb_{ss--}$ and $\coulc=\Coulomb_{ss+-}$ as defined earlier.
As discussed in Section \ref{sec:parityasym}, if the physical arrangement of \DWS\ and SET island as shown in \fig{fig:DoubleWellSystem} is symmetric about a line bisecting the double well potential, then $\coulc=0$.  Asymmetry results in $\coulc\neq0$.

Finally, we include the internal dynamics of the leads and of the  island, which is assumed to have at most a single electron, as well as tunnelling between the leads and the island.  The Hamiltonians for these parts of the complete system are given in \eqns{eqn:hisle} to (\ref{eqn:htun}). 
We now have the ingredients for the model Hamiltonian of the \DWS, SET island and leads
\begin{equation}
\htot=\hhub+\hat \Coulomb+\hisle+\hleads+\htun.\label{eqn:htotTP}
\end{equation}

\subsection{Measurement of symmetric configuration}

In this section, we will assume that the SET is placed symmetrically with respect to the \DWS, so $\coulc=0$. The triplet states and $\ket{\Dminus}$ are energy eigenstates of both $\hhub$ and $\hmeas$.  Similarly, since $\coulb$ and $\spenergydiff$ are small, $\ket{\singlet}$ is approximately an eigenstates of both $\hhub$ and $\hmeas$.  The induced charge on the SET island is different for the triplet states compared with the singlet, due to their slightly different charge configurations, so different currents flow through the SET depending on the subspace the electron pair is in.   Thus, distinguishing distinct currents through the SET yields  (approximately) QND projective measurements onto the singlet or triplet subspaces.  This is analogous to the situation described for the single particle in section \ref{sec:SP} for doing QND measurements in the parity eigenbasis.

\label{sec:TPheurmeas}

The induced shift in the SET island energy depends on the state of the \DWS.  
We can calculate the  SET island energy shift by imagining the \DWS\ in a given state,  then adiabatically turning on the SET--\DWS\ interaction, $\hat\Coulomb$. Physically, this corresponds to slowly bringing the occupied SET island close to the \DWS, and observing the change in energy of the total system during this process.  Comparing the adiabatic energy shift for a \DWS\ in a singlet with the energy shift for a \DWS\ in a triplet gives the differential shift of the SET island between the singlet and a triplet states.

A triplet state is an energy eigenstates of $\hhub+\alpha\hmeas$, where $0\leq\alpha\leq1$ is the adiabatic parameter controlling the coupling strength.   Therefore the adiabatic variation of the coupling does not change the eigenstate, just the eigenenergy,
\begin{equation}
\Delta E_{\tripup} =\bra{\tripup}(\hhub+\hmeas)\ket{\tripup}-\bra{\tripup}(\hhub)\ket{\tripup}=a.
\end{equation}

For a singlet state, which is almost, but not quite, an eigenstate of $\hhub+\alpha\hmeas$, we can estimate the induced shift by calculating the same adiabatic energy shift for the ground state $\ket{\tilde\singlet_\alpha}=\ket{\singlet}-\frac{\alpha\coulb+\spenergydiff}{U}\ket{\Dplus}$ which is very close to the singlet. We find
\begin{equation}
\Delta E_{\singlet} \approx\bra{\tilde\singlet_1}(\hhub+\hmeas)\ket{\tilde\singlet_1}-\bra{\tilde\singlet_0}\hhub\ket{\tilde\singlet_0}=a-\coulb(\coulb+2\spenergydiff)/U.
\end{equation}

Thus, the difference between the SET island energy for the triplet and singlet state is $\Delta E=\Delta E_T-\Delta E_S=\coulb(\coulb+2\spenergydiff)/U$, which corresponds to a differential induced SET island energy depending on the state of the \DWS.  It should therefore be possible to arrange the lead energies so that the SET current also depends on the state of the \DWS.  By tuning the lead chemical potentials so that $\mu_l>E_T>\mu_r>E_S$, current flows through the SET if the \DWS\ is in the triplet subspace, but does not for the subspace $\{\ket{\singlet},\ket{\Dplus}\}$.  This configuration is shown in \fig{fig:TwoParticleEnergyLevel}, where $E_T=\omega_{6,7,8}$ lies in between the lead Fermi energies, and $E_S=\omega_2$ lies below the Fermi energies.  The other levels shown in \fig{fig:TwoParticleEnergyLevel} represent possible inelastic transitions as lead electrons tunnel onto the SET island, and are described in more detail in appendix \ref{app:lindop}.  In this manner, the two subspaces may be distinguished by measuring the SET current.  

Since the measurement is QND in the triplet subspace, we use the same arguments as section \ref{sec:SPmeastime} to estimate the SET current when the \DWS\ is a triplet.  Assuming electrons tunnel between the leads and the SET island at a rate $\gamma$, then for the configuration of lead energies described above, the current for the triplet state will be $i_T=e\gamma/2$.  

The singlet state is approximately an eigenstate of $\hhub$ and $\hmeas$, so  the same reasoning concludes that the singlet current should be approximately zero, $i_S\approx0$.  The measurement time for distinguishing these two currents is then roughly $\tmeas=2/\gamma$, just as in section \ref{sec:SPmeastime}.

As mentioned above, the singlet state is not quite an eigenstate of the system or measurement Hamiltonians, so there are corrections to the latter part of this argument.  The dynamics mix the singly occupied state $\ket{\singlet}$ and the doubly occupied state $\ket{\Dplus}$.  Thus there is a small amplitude for the evolution to induce transitions from $\ket{\singlet}$ to $\ket{\Dplus}$.  In general these transitions are strongly inhibited since there is a large energy gap $\sim U$ to excite the \DWS\ to the doubly occupied state.  Appendix \ref{sec:singletmixing} shows that in the steady state, the probability for the \DWS\ to be in a singlet state is very close to unity, $\bra{
S}\rho^{ss}\ket{S}=\bra{
S}\rho^{ss}_0+\rho^{ss}_1\ket{S}=1-(\coulb+\spenergydiff)^2/\DoubleOccupationEnergy^2$. 
 This means that the measurement on the singlet subspace is indeed almost QND, 
 since the singlet is not changed greatly during measurement.  Associated with the infrequent fluctuations between $\ket{\singlet}$ and $\ket{\Dplus}$ is a small current.  To estimate an upper bound on the current, $i_\singlet$ that could flow through the SET when the \DWS\ is in the singlet state, we compute the rate at which electrons cycle on and off the SET island.  In appendix \ref{app:TPtraj} we show  $i_\singlet<2e\gamma\frac{ \spenergydiff^2{\left( \coulb  + \spenergydiff  \right) }^2}{\DoubleOccupationEnergy^4}\ll i_T$.  This shows that $i_S$ and $i_T$ are very different, and the measurement is indeed close to QND.

\subsection{Effect of asymmetry}

Asymmetry, $\coulc\neq0$, couples the states $\ket{\Dplus}$ and $\ket{\Dminus}$, evident in the form of $\hmeas$, in \eqn{eqn:hmeas}. We find the steady state probability for the \DWS\ to be in a singlet, $\bra{\singlet}\rho\ket{\singlet}=1-(\coulb+\spenergydiff)^2/(2\DoubleOccupationEnergy^2)$.  We also compute the rate at which electrons cycle on and off the SET island, described in appendices \ref{app:TPtraj} and \ref{app:assymBk}, which leads to an upper bound on the current given by $i_\singlet<e(\gamma_l+\gamma_r)\frac{ (\coulb^2+\spenergydiff^2){\left( \coulb  + \spenergydiff  \right) }^2}{2\DoubleOccupationEnergy^4}$.  Both the steady-state singlet probability, and the upper bound on the current for the asymmetric case are similar to the results for the symmetric case derived in appendices \ref{sec:singletmixing} and \ref{app:TPtraj}, indicating the measurement is rather \emph{insensitive} to SET asymmetry.   For larger values of $\epsilon$, when $\frac{2 x_\isle |\vec{r}_\isle^\mathrm{\,sym}|}{|\vec{r}_L-\vec{r}_\isle^\mathrm{\,sym}|^2}\gtrsim \inner{L}{R}$, the \DWS\ is driven into the doubly occupied subspace, which is the basis of an alternative singlet--triplet measurement scheme \cite{SeansPaper}.

\subsection{Summary}
 
In principle the scheme outlined above enables measurement  in the singlet-triplet basis.  We have shown that it is possible to measure distinguishable currents through the SET depending on the state of the \DWS.  The energy scales for the singlet--triplet measurement are smaller than those for the single-particle system, by a factor of $(\coulb+\spenergydiff)/\DoubleOccupationEnergy$, and this requires the lead temperatures to be smaller by a similar factor.  We have also established that asymmetry in the fabrication of the device is less of a problem for this measurement scheme than for the single particle scheme.  
 
\section{Estimation of experimental parameter values}\label{sec:estimationofparameters}

We now estimate the required parameters for various experimentally realisable systems.  Firstly we have assumed that $\coulb, \spenergydiff\ll \DoubleOccupationEnergy$.  We also require that the temperature be smaller than the SET energy shift induced by the \DWS. Thus, for the single particle case, we require that $\Boltz T\ll\coulb$,  and \fig{fig:TwoParticleEnergyLevel} indicates that for the two-particle system, the temperature must be smaller than the splitting between $\omega_2$ and $\omega_3$, i.e. $\Boltz T\ll {\coulb}(\coulb+2\spenergydiff)/{\DoubleOccupationEnergy}$.  This is obviously a tight constraint on the temperature of the system.  Furthermore, as discussed in section \ref{app:noncomME}, we require the line-width of the SET island state must be smaller than the energy level splittings, i.e.$\gamma\ll\coulb$ for the single-particle case and  $\gamma\ll{\coulb}(\coulb+2\spenergydiff)/{\DoubleOccupationEnergy}$ for the two-particle case.  Finally, cotunneling will contribute a background current due to tunneling via virtual population of the SET island.

In order to estimate the various parameters introduced for this problem, we need to estimate the overlap $\inner{L}{R}$.  We approximate  the localised states as $s$-orbitals bound to  each site, so that $\inner{\vec{r}}{L}=\InverseBohrRadius^\frac{3}{2}e^{-\InverseBohrRadius|\vec{r}-\vec{r}_L|}/\sqrt{\pi}$, where $\InverseBohrRadius$ is the inverse Bohr radius.  The integral $\inner{L}{R}$ may be performed in prolate ellipsoidal coordinates \cite{flu99} to give $\inner{L}{R}=(1+\InverseBohrRadius \DonorSep+\InverseBohrRadius^2 \DonorSep^2/3)e^{-\InverseBohrRadius \DonorSep}$, where $\DonorSep=|\vec{r}_L-\vec{r}_R|$ is the separation between the double well minima.

We can estimate $\spenergydiff$, the tunnelling rate between localised states, for the case of an $s$-orbital bound to a donor atom, and we will use this estimate for the case of electrostatically defined gates as well.  Following a similar argument to the derivation of \eqn{eqn:delta}, we can show that 
$\spenergydiff\approx2(\bra{R}\hat H\ket{L}-\inner{L}{R}\bra{L}\hat H\ket{L})$, 
where $\hat{H}=\hat{\vec{p}}^2/(2 m_e)+V_L(\vec{r})+V_R(\vec{r})$ is the time-independent, single-particle  Hamiltonian.  Assuming $V_{L, R}=q/|\vec{r}_{L, R}|$, we may again evaluate the integrals $\bra{R}\hat H\ket{L}$ and $\bra{L}\hat H\ket{L}$ in prolate ellipsoidal coordinates \cite{flu99} to find that the single particle tunnelling rate is given to reasonable approximation by $\spenergydiff\approx\frac{2}{3}{ q}{\InverseBohrRadius^2}\DonorSep \,e^{-\DonorSep \InverseBohrRadius} $.

For many materials, e.g. Si or GaAs, $\coulcoeff\approx 0.1$ eV nm.  For a donor atom system in Si, where $|\vec{r}_s|\approx10$ nm is a reasonable estimate for the height of the SET island above the donor system, and $\DonorSep\approx 30$ nm and a Bohr radius of $\mu^{-1}=3$ nm, giving $\inner{L}{R}=2\times10^{-3}$, and ignoring the anisotropic effective mass of Si  \cite{vri00a,bar03}, we have $\coulb\approx\coulcoeff\inner{L}{R}/|\vec{r}_s|\approx 20$ \micro{eV}.  Similarly $\spenergydiff\approx15$ \micro{}eV.  These figures could be increased to perhaps 200 \micro{}eV using an external $J$-gate, since the overlap integral depends exponentially on the $J$-gate potential.  Finally we estimate $\DoubleOccupationEnergy\sim10$ meV \cite{vri00a,bar03} for donor impurity systems.
In an electrostatically defined system such as GaAs dots, reasonable estimates for the various parameters are $\spenergydiff\approx150$ \micro{}eV, $\coulb\approx100$ \micro{}eV and $\DoubleOccupationEnergy\approx1$ mev \cite{bur99, bar03}.

For example, suppose $\DoubleOccupationEnergy\sim1$ meV and $\spenergydiff\approx\coulb\sim100$ \micro{}eV, so $\spenergydiff/\DoubleOccupationEnergy\sim0.1$ then for the two particle case we have $\Boltz T\ll 100$ \micro{}eV, i.e. $T\ll 1$ K.  Therefore, it is conceivable that the singlet--triplet measurement could be done at 0.3 K in electrostatically defined dots, which is an accessible electronic temperature.  In double donor systems, such as the Kane proposal $\spenergydiff\approx\coulb\sim100$ \micro{}eV is still reasonable, but since $\DoubleOccupationEnergy\sim10$ meV, the relevant temperature is around ten times smaller, which is probably too small to be practical.  This problem would be resolved if a sufficiently large $J$-gate voltage could be applied to increase $\spenergydiff$ and $\coulb$.   
Assuming that $\gamma_l=\gamma_r=\Boltz T=30 $ \micro{}eV $\approx 10^{11}\textrm{ s}^{-1}$, then $i_T\approx1$ nA.  The proposals are likely to work at temperatures $\Boltz T\sim\coulb,\spenergydiff$ as well, but with faster mixing times and longer measurement times.  The fundamental requirement for both of the proposals in this paper is that the overlap between the localised wavefunctions, $\inner{L}{R}$, be as large as possible, and preferably as large as about 0.1.

Finally, we estimate the effect of cotunneling by comparing the conductance due to resonant tunnelling processes, $\gres$, with that due to cotunneling, $\gcot$.  For weak coupling between the leads and the SET island these quantities are given by \cite{dat95,fur95}
\begin{equation}
\gres=\frac{G_L G_R}{G_L+G_R} \textrm{ and }
\gcot=\frac{\pi \hbar G_L G_R}{3e^2}\frac{(\Boltz T)^2}{\coulb^2},
\end{equation}
where $G_L$ and $G_R$ are the conductances of the left and right SET-lead tunnel barrier.  For the sake of estimation, we assume that these are equal to $G_L=G_R=\xi e^2/\hbar$ with $\xi\ll1$ for weak coupling.  The additional current due to cotunneling is small as long as $\gcot\ll\gres$, i.e.\ when $\Boltz T\ll\coulb/\sqrt{\xi}$, which is a less stringent constraint than above.  Therefore, as long as the previously discussed conditions are met, cotunnelling is small. 

\section{Conclusion}
 
 In this paper, we have presented and analysed a proposal for performing measurements in non-localised bases of both singly- and doubly-occupied double wells.
 
The physical mechanism by which the measurements operate is to detect the small variation in electronic charge density near the midpoint of the \DWS.  Based on reasonable estimates for the system parameters the small difference in the Coulomb potential at an SET island due to the different charge distributions of different states of the \DWS\ is in principle detectable.  The detected signal is the current through the SET island.

The main difficulties in these schemes is the precision with which the SET island must be placed at the mid-plane of the \DWS\ and the required operating temperatures.  For the single particle parity measurement, a misplaced SET island produces a measurement in the localised basis, which has been discussed in the past \cite{wis01}.  

The two-particle singlet-triplet measurement scheme is less sensitive to asymmetry in the placement of the SET, but requires very low temperatures to work effectively.

The required tolerance to such misplacement is at the edge of current fabrication technology of 1 nm or less, and may be achievable in light of recent experiments \cite{sch03}.  Other constraints such as operating temperature and tunnelling rates are experimentally achievable.

TMS thanks the Hackett committee, the CVCP and Fujitsu for financial support.  SDB acknowledges support from the E.U. NANOMAGIQC project (Contract no. IST-2001-33186).  HSG acknowledges financial support from Hewlett-Packard.


 
 \appendix
 
\section{Master equation for symmetric single particle system}\label{app:mesp}
 
The Hamiltonian for the complete system of double well, SET and leads is given by
\begin{equation}
\htot=\hsp+\hat \Coulomb+\hisle+\hleads+\htun.
\end{equation}

Following the derivation of  \citet{wis01}, we can write down a master equation for the reduced density matrix, $\rhos$, for the system consisting of the double well plus the SET island.  The dissipative terms are formally the same, where we identify $c_+$ and $c_\isle$ respectively with $c_1$ and $b$ in their notation.  The result is
\begin{eqnarray}
\dot \rhos(t)&=&{}-i \commute{\hsp+\hisle}{\rhos(t)}\nonumber\\
&&{}+\{\gamma_l(1- f_l(\omega_0))+\gamma_r (1-f_r(\omega_0))\}\lind[(1-\n_+)c_\isle]\rhos(t)\nonumber\\
&&{}+\{\gamma_l f_l(\omega_0)+\gamma_r f_r(\omega_0)\}\lind[(1-\n_+)c_\isle^\dagger]\rhos(t)\nonumber\\
&&{}+\{\gamma_l'(1-f_l(\omega_0+\coulb))+\gamma_r' (1-f_r(\omega_0+\coulb))\}\lind[\n_+c_\isle]\rhos(t)\nonumber\\
&&{}+\{\gamma_l' f_l(\omega_0+\coulb)+\gamma_r' f_r(\omega_0+\coulb)\}\lind[\n_+c_\isle^\dagger]\rhos(t),\label{eqn:SPMaster}
\end{eqnarray}
where $f_{l(r)}$ is the Fermi distribution for lead $l(r)$, $\lind[A]B\equiv \jump[A]B-\A[A]B\equiv ABA^\dagger-\frac{1}{2}(A^\dagger A B+B A^\dagger A)$ and $\gamma_i=\pi g_i |T_{ik_0}|^2$ and $\gamma_i'=\pi g_1 |T_{ik_0'}|^2$, where $g_i$ is the density of states in lead $i$, $k_0=\sqrt{2m \omega_0/\hbar}$ and $k_0'=\sqrt{2m (\omega_0+\chi)/\hbar}$.


We take $f_l(\omega_0)=f_l(\omega_0+\coulb)=1=f_r(\omega_0)$ and $f_r(\omega_0+\coulb)=0$, as shown in \fig{fig:SingleParticleEnergyLevel}, and then \eqn{eqn:SPMaster} becomes
\begin{equation}
\dot \rhos(t)={}-i \commute{\hsp+\hisle}{\rhos(t)}
+(\gamma_l+\gamma_r )\lind[\n_-^\dagger c_\isle^\dagger]\rhos(t)
+\gamma_r'\lind[\n_+c_\isle]\rhos(t)
+\gamma_l'\lind[\n_+^\dagger c_\isle^\dagger]\rhos(t).\label{eqn:appSPMaster1}
\end{equation}

 We will assume that the SET island is classical, in the sense that its reduced density matrix has no off diagonal terms.  This is justified since conservation of electron number between the leads and the island means that the electron number on the island is entangled with the electron number in the leads, and the lead degrees of freedom averaged over.  Therefore we write the double well plus SET island system in the separable form
\begin{equation}
\rhos=\rhodw_0\otimes\op{0}{0}+\rhodw_1\otimes\op{1}{1},\label{eqn:appR}
\end{equation}
where $\rhodw_{0(1)}$ is the state of the \DWS\ with $0(1)$ electrons on the SET island.  The reduced density matrix for the DWS alone is given by $\rho=Tr_s[R]=\rho_0+\rho_1$.


We now turn  the master equation into a pair of coupled equations for $\rho_0$ and $\rho_1$ by computing the matrix elements $\bra{0}\dot\rhos(t)\ket{0}$ and $\bra{1}\dot\rhos(t)\ket{1}$  using \eqn{eqn:SPMaster1}.
\begin{eqnarray}
\dot \rhodw_0(t)={}-i \commute{\hsp}{\rhodw_0(t)}
-(\gamma_l+\gamma_r )\A[\n_-]\rhodw_0(t)
+\gamma_r'\jump[\n_+]\rhodw_1(t)
-\gamma_l'\A[\n_+]\rhodw_0(t),\label{eqn:rho0}\\
\dot \rhodw_1(t)={}-i \commute{\hsp}{\rhodw_1(t)}
+(\gamma_l+\gamma_r )\jump[\n_-]\rhodw_0(t)
-\gamma_r'\A[\n_+]\rhodw_1(t)
+\gamma_l'\jump[\n_+]\rhodw_0(t).\label{eqn:rho1}
\end{eqnarray}
Since all the system operators in these equations are number operators, the equations are straightforward to solve.  We note that $\rhodw_{0,1}=\alpha_\pm \n_{\pm}$ are fixed points of the equations, for some coefficients $\alpha_\pm$ determined by rate balance.

Taking $\gamma_i=\gamma$, and with respect to the basis $\{\ket{+},\ket{-}\}$, the solution to these unconditional equations is 
\begin{eqnarray}
\rhodw_0(t)&=&\begin{bmatrix}
    \frac{1 + e^{-2\gamma t}}{2}\rhodw_0^{++}(0)+ \frac{1 - e^{-2\gamma t}}{2}\rhodw_1^{++}(0)   &  e^{-3\gamma  t/2}\rhodw_0^{+-}(0) \\
  e^{-3\gamma  t/2}\rhodw_0^{-+}(0)    &  e^{-2\gamma  t}\rhodw_0^{--}(0)
\end{bmatrix}\\
\rhodw_1(t)&=&\begin{bmatrix}
    \frac{1 - e^{-2\gamma t}}{2}\rhodw_0^{++}(0)+ \frac{1 + e^{-2\gamma t}}{2}\rhodw_1^{++}(0)   &  e^{-\gamma t/ 2}\rhodw_1^{+-}(0) \\
  e^{-\gamma t/2}\rhodw_1^{-+}(0)    &  (1-e^{-2\gamma  t})\rhodw_0^{--}(0)+\rhodw_1^{--}(0)
\end{bmatrix},\label{eqn:SPsymmsol}
\end{eqnarray}
where $\rhodw^{pq}=\bra{p}\rhodw\ket{q}$. 
 It is evident that the diagonal elements of the reduced density matrix for the \DWS, $\rhodw=\rhodw_0+\rhodw_1$, are constant as required for a QND measurement, whilst the off-diagonal elements decay with two characteristic time scales, the longest of which is $2/\gamma$, consistent with the measurement time computed assuming Poissonian tunnelling statistics in section \ref{sec:SPmeastime}.

\section{Approximate expressions for $\coulc$ and $\coulb$}\label{app:approximations}

We now estimate $\coulc$ using \eqn{eqn:coulint}.
For the purposes of this estimate, we will assume the SET island wavefunction is a delta function located at $\vec{r}_\isle=\{x_\isle,y_\isle,z_\isle\}$ ($x_\isle$ is the axial position of the island, and $x_\isle=0$ for a symmetric arrangement) as shown in \fig{fig:AsymmetricSystem}, i.e. $|\phi_\isle(\vec{r})|^2=\diracdelta(\vec{r}-\vec{r_\isle})$, so 
\begin{eqnarray}
\coulc&=&\Coulomb_{\isle\isle+-}=\coulcoeff\int \ud^3\vec{r}  \frac{\phi_+^*(\vec{r})\phi_-(\vec{r})}{|\vec{r}-\vec{r}_\isle|}= \bra{+}\frac{1}{|\vec{r}-\vec{r}_\isle|}\ket{-}\nonumber\\
&=&\coulcoeff\left(\bra{L}\frac{1}{|\vec{r}-\vec{r}_\isle|}\ket{L}-\bra{R}\frac{1}{|\vec{r}-\vec{r}_\isle|}\ket{R}+\bra{R}\frac{1}{|\vec{r}-\vec{r}_\isle|}\ket{L}-\bra{L}\frac{1}{|\vec{r}-\vec{r}_\isle|}\ket{R}\right),
\end{eqnarray}
where $\inner{\vec{r}}{L}=\phi_L(\vec{r})=\phi(\vec{r}-\vec{r_L})$, $\phi(\vec{r})$ is the localised wavefunction of a single site and $\vec{r_L}$ is the `centre-of-mass' of the left well.  
The last two terms cancel for all $x_\isle$, and if $x_\isle=0$ then the first two terms cancel also, hence our previously stated result that $\coulc=0$ for a symmetric configuration.  For $x_\isle\neq0$ we have
\begin{equation}
\bra{L}\frac{1}{|\vec{r}-\vec{r}_\isle|}\ket{L}=\int \ud^3\vec{r}  \frac{|\phi(\vec{r}-\vec{r_L})|^2}{|\vec{r}-\vec{r}_\isle|}=\int \ud x \ud y \ud z  \frac{|\phi(\{x-x_L,y,z\})|^2}{\sqrt{(x-x_\isle)^2+(y-y_\isle)^2+(z-z_\isle)^2}}.
\end{equation}
We assume the asymmetry is small so that $x_\isle$ is a small quantity, and we expand the square-root in a power series about $x_\isle=0$ to find 
\begin{equation}
\bra{L}\frac{1}{|\vec{r}-\vec{r}_\isle|}\ket{L}=\bra{L}\frac{1}{|\vec{r}-\vec{r}_\isle^\mathrm{\,sym}|}\ket{L}
+x_\isle\eta+O(x_\isle^2),
\end{equation}
where $\vec{r}_\isle^\mathrm{\,sym}$ is the intended, symmetric location of the SET island and $\eta=\int \ud^3\vec{r}  \frac{|\phi(\vec{r}-\vec{r_L})|^2 x}{|\vec{r}-\vec{r}_\isle^\mathrm{\,sym}|^3}$.  Following the same reasoning, we can show that 
\begin{equation}
\bra{R}\frac{1}{|\vec{r}-\vec{r}_\isle|}\ket{R}=\bra{R}\frac{1}{|\vec{r}-\vec{r}_\isle^\mathrm{\,sym}|}\ket{R}
-x_\isle\eta.
\end{equation}
We can estimate $\eta$ by assuming that the localised wavefunction is very tightly bound, so that $|\phi(\vec{r})|^2=\diracdelta(\vec{r})$, and then $\eta=\frac{x_L}{|\vec{r_L}-\vec{r}_\isle^\mathrm{\,sym}|^3}$.
Since $\bra{R}\frac{1}{|\vec{r}-\vec{r}_\isle^\mathrm{\,sym}|}\ket{R}=\bra{L}\frac{1}{|\vec{r}-\vec{r}_\isle^\mathrm{\,sym}|}\ket{L}$, we find that 
\begin{equation}
\coulc\approx \coulcoeff\frac{2 x_\isle x_L}{|\vec{r_L}-\vec{r}_\isle^\mathrm{\,sym}|^3}\lesssim \coulcoeff\frac{2 x_\isle}{|\vec{r_L}-\vec{r}_\isle^\mathrm{\,sym}|^2}.
\end{equation}

Estimating $\coulb$ is more difficult, and without detailed knowledge of the localised wavefunction $\phi(\vec{r})$ our estimate of it is somewhat less controlled than that of $\coulc$.  For a symmetric system we have
\begin{eqnarray}
\coulb&=&\Coulomb_{\isle\isle ++}-\Coulomb_{\isle\isle --}=\bra{+}\hat\Coulomb\ket{+}-\bra{-}\hat\Coulomb\ket{-}\nonumber\\
&=&\frac{1}{2+2\inner{L}{R}}(\bra{L}+\bra{R})\hat\Coulomb(\ket{L}+\ket{R})-\frac{1}{2-2\inner{L}{R}}(\bra{L}-\bra{R})\hat\Coulomb(\ket{L}-\ket{R})\nonumber\\
&=&2\bra{R}\hat \Coulomb\ket{L}-2\inner{L}{R}\bra{L}\hat\Coulomb\ket{L}+O(\inner{L}{R}^2).\label{eqn:delta}
\end{eqnarray}
Using the approximation that $\phi$ is tightly bound allows us to approximate
$
\bra{L}\hat\Coulomb\ket{L}\approx\frac{\coulcoeff}{|\vec{r}_L-\vec{r}_\isle^\mathrm{\,sym}|}.\label{eqn:deltaapprox}
$
We can estimate an upper bound on $\bra{R}\hat \Coulomb\ket{L}$ by considering that $\phi^*_L(\vec{r})\phi_R(\vec{r})$ is peaked with a maximum at the midpoint of the double well.  Thus, the potential at the island due to the charge distribution $\phi^*_L(\vec{r})\phi_R(\vec{r})$  will be less than the potential due to  the entire weight of this product located at the midpoint.  That is   
$
\bra{R}\hat \Coulomb\ket{L}\lesssim\frac{\coulcoeff\inner{L}{R}}{|\vec{r}_\isle^\mathrm{\,sym}|}
$. Therefore an estimate for the magnitude of $\coulb$ is
\begin{equation}
\coulb\approx \coulcoeff\inner{L}{R}\left(\frac{1}{|\vec{r}_\isle^\mathrm{\,sym}|}-\frac{1}{|\vec{r}_L-\vec{r}_\isle^\mathrm{\,sym}|}\right).\label{eqn:deltaestimate}
\end{equation}

\section{Derivation of master equation for non-commuting system and measurement Hamiltonians}\label{app:noncomME}

In this appendix we derive a master equation for a device whose system Hamiltonian, $\hsys$, does not commute with the measurement Hamiltonian, $\hmeas$.   The results from this appendix are used in Appendices \ref{app:asymmesp} and \ref{app:mesymdodws}, wherein $\hsys=\hhub$.  The total Hamiltonian for the device is taken to be 
\begin{equation}
\htot=\hsys+\hat \Coulomb+\hisle+\hleads+\htun,
\end{equation}
where $\hat \Coulomb=\n_\isle\otimes\hmeas$ and
$
\hisle=\omega_0 \n_\isle$,
$\hleads=\sum_k \omega_k(\n_{l k}+\n_{r k})$ and 
$\htun=\sum_k T_{l k} c^\dagger_{l k} c_\isle+T_{r k} c^\dagger_{r k} c_\isle+\hc
$, as given in section \ref{sec:system}.

The general method for this derivation follows several steps.  
\begin{enumerate}
\item We move to an interaction picture to transform away all the free dynamics. 
 
\item Using the Zassenhaus relation we factor the interaction Hamiltonian into a product of lead operators and a finite-dimensional operator acting on the \DWS\ and  SET.

\item By tracing over the lead modes, we derive a Markovian master equation for the \DWS\ and  SET density matrix, $R$.   In this master equation, the Fourier components of $B_k(t)$ appear in the Lindblad terms.
\end{enumerate}


We transform to an interaction picture with respect to the Hamiltonian $\hrot=\hsys+\hat \Coulomb+\hisle+\hleads$, so that $\htot=\hrot+\htun$, and the interaction picture Hamiltonian is  $\hi(t)=e^{i \hrot t}\htot e^{-i \hrot t}-\hrot=e^{i \hrot t}\htun e^{-i \hrot t}$.  In order to compute $\hi$ we first note that $\hleads$ and  $\hisle$  commute with all other terms in $\hrot$ so $e^{i \hrot t}=e^{i(\hsys+\hat \Coulomb) t}e^{i \hisle t}e^{i \hleads t}$.  Using the operator identities 
\begin{equation}
e^{x \n}c=c \textrm{ and } c e^{x \n}=e^x c,\label{eqn:opident}
\end{equation}
where $\n=c^\dagger c$ and $\commute{x}{c}=0$, we find that 
\begin{equation}
\hi(t)=\sum_k (T_{l k} c^\dagger_{l k}+T_{r k} c^\dagger_{r k})
e^{i(\omega_k-\omega_0)t} e^{i(\hsys+\hat \Coulomb) t}c_\isle e^{-i(\hsys+\hat \Coulomb) t} +\hc.
\end{equation}
Since $\commute{\hsys}{\hat\Coulomb}=\n_\isle\otimes\commute{\hsys}{\hmeas}\neq0$ the operator exponentials  appearing above cannot be factorised.  However we may simplify the expression using the Zassenhaus operator relation \cite{sri02}, which is an inverse relation to the Baker-Campbell-Hausdorff formula, and it states that \mbox{$e^{A+B}=e^A e^B \prod_{j\uparrow} e^{C_j[A,B]}$} where each $C_j[A,B]=(-1)^jC_j[B,A]$ is a sum of nested commutators, each term of which has $A$ and $B$ appearing at least once (e.g. $C_1[A,B]=-\frac{1}{2}\commute{A}{B}$ and $C_2[A,B]=-\frac{1}{6}(\commute{A}{\commute{A}{B}}-\commute{B}{\commute{B}{A}})$.  The $\uparrow$ in the index of the product indicates that the product is ordered in increasing order of $j$, since the factors in the product don't commute.  For our purposes, the detailed form of $C_j$ is not important.  

Taking $A=i \hsys t$ and $B=i\hat\Coulomb t=i \n_\isle \hmeas t$, and noting that $\n_\isle^2=\n_\isle$ so that  $C_j[i \hsys t,i \n_\isle \hmeas t]=(i t)^{j+1} \n_\isle C_j[\hsys,\hmeas]$ we have
\begin{eqnarray}
&&e^{i(\hsys+\hat \Coulomb) t}c_\isle e^{-i(\hsys+\hat \Coulomb) t}\nonumber\\
&=&e^{i \hsys t} e^{i \n_\isle \hmeas t} \prod_{j\uparrow} e^{ (i t)^{j+1} \n_\isle C_j[ \hsys , \hmeas]}c_\isle\prod_{j\downarrow} e^{- (i t)^{j+1} \n_\isle C_j[ \hsys , \hmeas]}e^{-i \n_\isle \hmeas t}e^{-i \hsys t},\nonumber\\
&=&e^{i \hsys t}  \prod_{j\downarrow} e^{- (i t)^{j+1} C_j[ \hsys , \hmeas]}e^{-i  \hmeas t}e^{-i \hsys t}c_\isle,\nonumber\\
&=&e^{i \hsys t} e^{-i(\hsys+\hmeas) t}c_\isle,
\end{eqnarray}
where the first equality follows from direct substitution into the Zassenhaus relation, the second equality follows from repeated applications of \eqns{eqn:opident}, and the final equality follows by inverting the Zassenhaus relation.  We may therefore write the interaction Hamiltonian as
\begin{eqnarray}
\hi(t)&=&\sum_k (T_{l k} c^\dagger_{l k}+T_{r k} c^\dagger_{r k})
e^{i(\omega_k-\omega_0)t} e^{i \hsys t} e^{-i(\hsys+\hmeas) t}c_\isle+\hc,\nonumber\\
&\equiv&\sum_k (T_{l k} c^\dagger_{l k}+T_{r k} c^\dagger_{r k})B_k(t) c_\isle+\hc,\\
\textrm{where }B_k(t)&=&e^{i(\omega_k-\omega_0)t} e^{i \hsys t} e^{-i(\hsys+\hmeas) t}\label{eqn:Bk}
\end{eqnarray}
 is an operator acting on the \DWS\ alone.

The state matrix $W$ of the entire closed system including the double well, SET island and leads evolves according to the Schr\"odinger equation in the interaction picture, taken to second order \cite{wis01}
\begin{equation}
W(t+\Delta t)=W(t)-i \Delta t\commute{\hi(t)}{W(t)}-\Delta t\int_t^{t+\Delta t}\ud \dummy \commute{\hi(t)}{\commute{\hi(\dummy)}{W(\dummy)}}.
\end{equation}
Making the first Markov approximation, we assume that the system may at any time be written as $W(t)=\rhoi(t)\otimes \rho_l\otimes \rho_r$, so that each lead is always in a thermal state.  Then averaging over lead degrees of freedom $\braket{\cdot}_{l,r}$, and noting that $\braket{c_{l(r)k}}=\braket{c^\dagger_{l(r)k}}=0$,  $\braket{c_{l(r)k} c_{r(l)k'}}=\braket{c^\dagger_{l(r)k} c_{r(l)k'}}=\braket{c^\dagger_{l(r)k} c^\dagger_{r(l)k'}}=0$ and $\braket{c^\dagger_{l(r)k} c_{l(r)k'}}=\diracdelta(k-k')f_{l(r)}(\omega_k)$, where $f_{l(r)}$ is the Fermi distribution for lead $l(r)$ and $\diracdelta(x)$ is the Dirac-delta function.  Then the SET island plus \DWS\ interaction picture density matrix, $\rhoi(t)$, satisfies
\begin{eqnarray}
\dot \rhoi(t)&=&-\int\ud \omega_k \left(g_\leadS |T_{\leadS k}|^2 f_\leadS(\omega_k)+g_\leadD |T_{\leadD k}|^2
f_\leadD(\omega_k)\right)\times\nonumber\\
&&\quad\quad\int_{-\infty}^t \ud \dummy\left\{\vphantom{B^\dagger_k}\nonumber\right.B_k(t)B^\dagger_k(\dummy)c_\isle c_\isle^\dagger \rhoi(\dummy)-B^\dagger_k(t) c_\isle^\dagger \rhoi(\dummy) c_\isle B_k(\dummy)\\
&&\hphantom{\quad\quad\int_{-\infty}^t \ud \dummy\left\{\right.}\left.-B^\dagger_k(\dummy) c_\isle^\dagger \rhoi(\dummy) c_\isle B_k(t)+\rhoi(\dummy)c_\isle c_\isle^\dagger B_k(\dummy)B^\dagger_k(t)\right\}\nonumber\\
&&-\int\ud \omega_k \left(g_\leadS |T_{\leadS k}|^2 (1-f_\leadS(\omega_k))+g_\leadD |T_{\leadD k}|^2 (1-f_\leadD(\omega_k))\right)\times\nonumber\\
&&\quad\quad\int_{-\infty}^t \ud \dummy\left\{\vphantom{B^\dagger_k}\nonumber\right.\left.B^\dagger_k(t)B_k(\dummy)c^\dagger_\isle c_\isle \rhoi(\dummy)-B_k(t) c_\isle \rhoi(\dummy) c^\dagger_\isle B^\dagger_k(\dummy)\right.\nonumber\\
&&\hphantom{\quad\quad\int_{-\infty}^t \ud \dummy\left\{\right.}\left.-B_k(\dummy) c_\isle \rhoi(\dummy) c^\dagger_\isle B^\dagger_k(t)+\rhoi(\dummy)c^\dagger_\isle c_\isle B^\dagger_k(\dummy)B_k(t)\right\}, \label{eqn:ReducedIntegral}
\end{eqnarray} where $g_i$ is the density of states for lead $i$.
We further assume that the dynamics of the system is slow compared to tunnelling rates etc. so that we may make the replacement $\rhoi(\dummy)\rightarrow\rhoi(t)$ in the above integrals, making \eqn{eqn:ReducedIntegral} local in time.  Equation (\ref{eqn:ReducedIntegral}) no longer depends on the lead degrees of freedom, and so is an equation for a finite dimensional system.  With the aid of some further approximations, we may perform the integrations over $\omega_k$ and $\dummy$, which we now do.

In order to do the integrals, we note that each term in \eqn{eqn:ReducedIntegral} is finite dimensional so has a finite dimensional matrix representation.  Further, we may write $B_k(t)$ as a discrete Fourier decomposition
\begin{equation}
B_k(t)=\sum_{m=1}^N e^{i(\omega_k-\omega_m)t} P_m\label{eqn:Bkfourier}
\end{equation}
for some finite $N$ and operators $P_m$.  From \eqn{eqn:Bk}, the explicit form of $P_m$ depends on the explicit form of $\hsys$ and is important for the discussion of the dynamics of the system. The operators $P_m$ for the single-particle DWS are given in \eqn{eqn:asymbk}. For the two-electron DWS, they are given in \eqn {eqn:FourierDecompBsym} and \eqn{eqn:FourierDecompBasym} for symmetric and asymmetric cases, respectively.  We now perform the integrations for one of the terms in \eqn{eqn:ReducedIntegral} as an example, to show explicitly the approximations we make.  For instance the second term of \eqn{eqn:ReducedIntegral}  is
\begin{eqnarray}
\int \ud\omega_k g_\leadS |T_{\leadS k}|^2 f_l(\omega_k)&&\int_{-\infty}^t\ud \dummy B^\dagger_k(t) c_\isle^\dagger \rhoi(t) c_\isle B_k(\dummy)\nonumber\\
=&&\sum_{mn}\int \ud\omega_k g_\leadS |T_{\leadS k}|^2 f_l(\omega_k)\int_{-\infty}^t\ud \dummy e^{i(\omega_k-\omega_m)t}e^{-i(\omega_k-\omega_n)\dummy}P^\dagger_m c_\isle^\dagger \rhoi(t) c_\isle P_n,\nonumber\\
=&&\sum_{mn}\int \ud\omega_k \gamma_\leadS |T_{\leadS k}|^2 f_l(\omega_k) \diracdelta(\omega_k-\omega_n) e^{i(\omega_n-\omega_m)t}P^\dagger_m c_\isle^\dagger \rhoi(t) c_\isle P_n,\nonumber\\
\approx&&\sum_{m} \gamma_l f_l(\omega_m) P^\dagger_m c_\isle^\dagger \rhoi(t) c_\isle P_m,
\end{eqnarray}
where $\gamma_l$ is defined in appendix \ref{app:mesp}.  The first equality follows from substituting the Fourier decomposition of $B_k(t)$, the second equality follows from evaluating the integral  over $\dummy$, and to make the final (approximate) equality we have made a rotating-wave approximation, where we take $e^{i(\omega_n-\omega_m)t}=\diracdelta_{m,n}$.  This is reasonable if the frequency difference $\omega_n-\omega_m$ (for $n\neq m$) is sufficiently large, since when we come to solve the resulting differential equation terms containing a factor $e^{i(\omega_n-\omega_m)t}$ will be rotating rapidly, and so average to zero, to good approximation.  This approximation is reasonable when $\gamma_{l(r)}$ is much smaller than the smallest energy level separations, $\gamma_{l(r)}\ll\omega_n-\omega_m$, for $n\neq m$.

Applying these arguments to the other terms in \eqn{eqn:ReducedIntegral} results in the master equation for $\rhoi(t)$
\begin{eqnarray}
\dot \rhoi(t)&=&\{\sum_m\left(\gamma_l f_l(\omega_m)+\gamma_r f_r(\omega_m)\right)\lind[P_m^\dagger c^\dagger_\isle]\rhoi(t)\nonumber\\
&&+\left(\gamma_l (1-f_l(\omega_m))+\gamma_r (1-f_r(\omega_m))\right)\lind[P_m c_\isle]\rhoi(t)\},\label{eqn:Interaction2PMasterEquation}
\end{eqnarray}
where again $\lind[A]B\equiv ABA^\dagger-\frac{1}{2}(A^\dagger A B+B A^\dagger A)$.
This forms a generalisation of the results of Wiseman \etal\ \cite{wis01} to the situation where the measurement Hamiltonian ($H_{CB}$ in their notation) does not commute with the free Hamiltonian of the system ($H_0$ in their notation).  Equation (\ref{eqn:Interaction2PMasterEquation}) shows the importance of the Fourier decomposition of the system operator $B_k(t)$ -- the Fourier components of $B_k(t)$, and their adjoint, form the Lindblad operators in the master equation, and it is through these components that the \DWS\ interacts with the SET island.

Returning to the Schr\"odinger picture, the master equation is given by 
\begin{eqnarray}
\dot \rhos(t)=-i\commute{\hrot}{\rhos}&+&\sum_m\{\left(\gamma_l f_l(\omega_m)+\gamma_r f_r(\omega_m)\right)\lind[P_m^\dagger c^\dagger_\isle]\rhos(t)\nonumber\\
&+&\left(\gamma_l (1-f_l(\omega_m))+\gamma_r (1-f_r(\omega_m))\right)\lind[P_m c_\isle]\rhos(t)\},\label{eqn:2PMasterEquation}
\end{eqnarray}

\section{Master equation for asymmetric single particle system}\label{app:asymmesp}

We now derive a master equation for the single-particle \DWS\ for the case that the SET island is not placed symmetrically. 
Equation (\ref{eqn:asympot}) gives the Hamiltonian for the Coulomb interaction between the SET island and \DWS\ as 
\begin{equation}
\hat \Coulomb=\n_\isle\otimes(\coulb\n_++\coulc(c_+^\dagger c_- + c_-^\dagger c_+)\equiv\n_\isle\otimes\hmeas.
\end{equation}
When $\coulc\neq0$, $\commute{\hmeas}{\hsp}\neq0$, so we use the derivation of the master equation in appendix \ref{app:noncomME}.

Firstly, from \eqn{eqn:Bk}, 
\begin{eqnarray}
B_k(t)&=&e^{i(\omega_k-\omega_0)t} e^{i \hsp t} e^{-i(\hsp+\hmeas) t},\\
&=&\sum_{m=1}^4 e^{i(\omega_k-\omega_m)t} P_m,\\
&=&e^{i \omega_k t}\left(
e^{-i\omega_0 t}\tilde n_-
+e^{-i(\omega_0+\coulb+\Delta) t}\tilde\sigma_-
+e^{-i(\omega_0+\coulb) t}\tilde n_+
+e^{-i(\omega_0-\Delta) t}\tilde\sigma_+
\right)\label{eqn:asymbk}
\end{eqnarray}
where $\tilde\sigma_+=-\tilde\sigma_-^\dagger= -\frac{\coulc}{\coulb+\spenergydiff}\ket{+}\bra{-}$ and $\tilde n_\pm=\n_\pm -\tilde\sigma_\pm$.  Thus, the Fourier components, $P_m$, of $B_k(t)$ are the operators appearing in \eqn{eqn:asymbk} associated with the four Fourier frequencies $\omega_m\in\{\omega_0, \omega_0+\coulb+\spenergydiff,\omega_0+\coulb,\omega_0-\spenergydiff\}$.  Here we have neglected terms of $O(\coulc^2)$ or higher, since these are negligible.

The master equation for the SET and \DWS\ is then
\begin{eqnarray}
\dot \rhos(t)&=&-i \commute{\hsp+\hisle}{\rhos(t)}
+(\gamma_l+\gamma_r )\lind[\tilde n_-c_\isle^\dagger]\rhos(t)
+\gamma_r'\lind[\tilde n_+c_\isle]\rhos(t)
+\gamma_l'\lind[\tilde n_+^\dagger c_\isle^\dagger]\rhos(t)\nonumber\\
&&\hphantom{{}-i \commute{\hsp}{\rhos(t)}}
+(\gamma_l+\gamma_r )\lind[\tilde\sigma_+^\dagger c_\isle^\dagger]\rhos(t)
+\gamma_r'\lind[\tilde\sigma_-c_\isle]\rhos(t)
+\gamma_l'\lind[\tilde\sigma_-^\dagger c_\isle^\dagger]\rhos(t).\label{eqn:asymSPME}
\end{eqnarray}
This expression agrees with \eqn{eqn:SPMaster1} in the limit that $\coulc\rightarrow0$.  We again assume the SET island does not maintain coherence, as expressed in \eqn{eqn:appR}, and we then can solve \eqn{eqn:asymSPME} for $\rhodw_0(t)$ and $\rhodw_1(t)$.

The most important quantity to derive from this master equation is the mixing time.  By taking the Laplace transform of \eqn{eqn:asymSPME}, we find poles at $0, \frac{-3\gamma {\coulc }^2}{2\left( \coulb  + \spenergydiff  \right) ^2},-\gamma/2,-3\gamma/2$ and $-2 \gamma$.  All but the second of these poles appear as rates in the solution for the symmetric case, \eqn{eqn:SPsymmsol}.  The second pole is very small, and corresponds to the mixing rate induced by the asymmetry in the SET island placement.  For times $t\ll\tmix=\frac{2\left( \coulb  + \spenergydiff  \right) ^2}{3\gamma {\coulc }^2}$, the solution to the master equation is essentially the same as \eqn{eqn:SPsymmsol}.  On a time-scale $t\sim\tmix$, the diagonal elements also decay, so that the steady-state solution in the ordered basis $\{\ket{+},\ket{-}\}$ is
\begin{equation}
\rhodw_0=\begin{bmatrix}\frac{{\coulc }^4}{2{\left( \coulb  + \spenergydiff  \right) }^4} & 0 \\0 & \frac{{\coulc }^4}{2{\left( \coulb  + \spenergydiff  \right) }^4}\end{bmatrix}
\textrm{ and }
\rhodw_1=\begin{bmatrix}\frac{{\coulc }^2}{{\left( \coulb  + \spenergydiff  \right) }^2} & -\frac{{\coulc }}{{\left( \coulb  + \spenergydiff  \right) }}  \\-\frac{{\coulc }}{{\left( \coulb  + \spenergydiff  \right) }}  & 1-\frac{{\coulc }^2}{{\left( \coulb  + \spenergydiff  \right) }^2}
\end{bmatrix},\label{eqn:sssolsp}
\end{equation}
where we have kept only the highest order term in $\coulc$ for each matrix element.  This steady-state solution corresponds to the \DWS\  being (almost) in its ground state, $\ket{-}$, with the SET in the closed state, occupied by a single electron.

\section{Master equation for a doubly occupied \DWS}\label{app:mesymdodws}

In this appendix we derive a master equation for the dynamics of the singlet-triplet measurement scheme.  Initially we consider a symmetrically placed SET island.  We first derive the Lindblad operators that appear in the master equation, and give a physical interpretation to the discrete spectrum, $\{\omega_m\}$, that appears in their derivation.  We then give a quantum trajectories analysis of the measured currents in the triplet subspaces and singlet subspaces.  Next we compute the degree of mixing in the singlet subspace induced by the fact that the singlet is not an eigenstate of the dynamics.  Finally, we provide a derivation of the jump operators for an asymmetrically place SET.

\subsection{Lindblad operators}\label{app:lindop}

The device Hamiltonian is given by \eqn{eqn:htotTP} so we can use the results of appendix \ref{app:noncomME} to derive a master equation for the dynamics of the \DWS\ and SET island coupled to the leads.  The crucial quantity to evaluate is  $B_k(t)$, whose definition is given in \eqn{eqn:Bk}.  Then identifying the Fourier components, $P_m$ of $B_k(t)$, as in \eqn{eqn:Bkfourier}, provides the operators that appear in the Lindblad terms of the master equation, \eqn{eqn:Interaction2PMasterEquation}.

We assume that $\DoubleOccupationEnergy$ and $\coula$ are relatively large energies, whilst $\coulc$, $\coulb$ and $\spenergydiff$ are relatively small. In fact, for the symmetric case, $\coulc=0$, and we will investigate this `ideal' situation first.  Clearly, if $\coulc=0$, then the dynamics is even more restricted, so that the measurement Hamiltonian, $\hat\Coulomb=\n_\isle\otimes\hmeas$, only couples the spatially symmetric states $\ket{\singlet}$ and $\ket{\Dplus}$, (see \eqn{eqn:hmeas}), so we may restrict our analysis to the 2-dimensional subspace spanned by the ordered basis $\{\ket{\singlet},\ket{\Dplus}\}$.  In this restricted basis we may decompose $B_k(t)$ into four Fourier components
\begin{equation}
B_k^{\{\singlet,\Dplus\}}(t)=e^{i\omega_k t}\Big(e^{i\DoubleOccupationEnergy
t}\begin{bmatrix}
    0  & 0 \\
    -\frac{ \coulb}{\DoubleOccupationEnergy}   &  0
\end{bmatrix}+
e^{i\frac{\coulb^2+2 \coulb\spenergydiff}{\DoubleOccupationEnergy}t}
\begin{bmatrix}
    1  & -\frac{ \coulb+\spenergydiff}{\DoubleOccupationEnergy}  \\
    -\frac{ \spenergydiff}{\DoubleOccupationEnergy} &  0
\end{bmatrix}
+e^{-i\frac{\coulb^2+2 \coulb\spenergydiff}{\DoubleOccupationEnergy}t}\begin{bmatrix}
    0  & \frac{ \spenergydiff}{\DoubleOccupationEnergy}  \\
    \frac{ \coulb+\spenergydiff}{\DoubleOccupationEnergy} &  1
    \end{bmatrix}
    +e^{-i\DoubleOccupationEnergy t}
    \begin{bmatrix}
    0  & \frac{ \coulb}{\DoubleOccupationEnergy} \\
    0   &  0
\end{bmatrix}\Big),\label{eqn:FourierDecompBsym}
\end{equation}
where for simplicity (and without loss of generality) we have set $\coula=-\omega_0$ so that a common overall factor of $e^{i(\coula+\omega_0)t}$ conveniently vanishes.  Also, we have discarded terms of order $\coulb^2, \coulb\spenergydiff$ and $\spenergydiff^2$ appearing in the matrices, since these are small, but they are retained in exponents where they are the lowest order terms that lift the degeneracy of the $\omega_m$.  We will refer to the operators appearing in \eqn{eqn:FourierDecompBsym} as $P_1, P_2, P_4$ and $P_5$ respectively.  
For completeness,
\begin{equation}
B_k(t)=B_k^{\{\singlet,\Dplus\}}(t)+e^{i \omega_k t}
\left(P_3+P_6+P_7+P_8\right),\label{eqn:BkTPsym}
\end{equation}
where $P_3=\op{\Dminus}{\Dminus}$, $P_6=\op{\tripup}{\tripup}$, $P_7=\op{\tripud}{\tripud}$ and $P_8=\op{\tripdown}{\tripdown}$. 
This decomposition of $B_k(t)$ shows that there are eight Lindblad operators, $P_1,...,P_8$.

The frequencies $\omega_m$, associated with the measurement process, are given by $\omega_{3,6,7,8}=0$, $\omega_4=-\omega_2=\frac{\coulb^2+2\coulb\spenergydiff}{\DoubleOccupationEnergy}$ and  $\omega_5=-\omega_1=
 U$ (where, again, we have set $\coula=-\omega_0$, otherwise we have an overall offset of $\coula+\omega_0$ to our energy scale).  These energies are shown relative to the  lead chemical potentials, $\mu_l$ and $\mu_r$, in \fig{fig:TwoParticleEnergyLevel}.  This choice determines the coefficients of the Fermi factors in the master equation, \eqn{eqn:Interaction2PMasterEquation}. 

 We interpret the energies $\hbar \omega_m$ as the change in energy of electrons tunnelling between the SET island and a lead.  Thus, since $\omega_{3,6,7,8}=0$, the corresponding processes, $P_{3,6,7,8}$ are associated with elastic tunnelling between the lead and the SET.  This can only induce dephasing of the \DWS, since no energy is exchanged between the leads and the \DWS.  These elastic processes therefore do not induce mixing in the \DWS, and are the origin of the QND projective nature of the measurement in the triplet subspace.
 
 Conversely, $\omega_{1,2}<0$, so $P_{1,2}$ correspond to inelastic lead--SET tunnelling processes which \emph{gain} an energy $\hbar \omega_{1,2}$.  This additional energy in the lead is provided by the electron-pair in the \DWS\ which \emph{loses} energy.    Similarly, processes $P_{4,5}$ correspond to lead electrons losing energy as the \DWS\ becomes excited.  We therefore expect that there will be some measurement-induced energy relaxation associated with the measurement of a singlet state. 
 
We note in passing that these elastic and inelastic processes in the detector have counterparts in the  measurement of a \DWS\ by a point contact detector, as described in \cite{stace:136802}.

\subsection{Master equation in singlet subspace}\label{sec:singletmixing}

The singlet and triplet subspaces are not mixed at all by the dynamics, so we derive a master equation for the state of the \DWS\ and SET in the singlet subspace, $R=\rhodw_0\otimes\op{0}{0}+\rhodw_1\otimes\op{1}{1}$, using the results of appendix \ref{app:noncomME}.  In particular, the $P_m$ that appear in \eqn{eqn:FourierDecompBsym} form the Lindblad operators in \eqn{eqn:Interaction2PMasterEquation} and the $\omega_m$ appear as arguments to the Fermi functions in \eqn{eqn:Interaction2PMasterEquation},
\begin{eqnarray}
\dot\rhodw_0&=&-i\commute{\hhub}{\rhodw_0}+\gamma'(\jump[P_4]\rhodw_1+\jump[P_5]\rhodw_1-\A[P_1^\dagger]\rhodw_0-\A[P_2^\dagger]\rhodw_0),\nonumber\\
\dot\rhodw_1&=&-i\commute{\hhub+\hmeas}{\rhodw_1}-\gamma'(\A[P_4]\rhodw_1+\A[P_5]\rhodw_1-\jump[P_1^\dagger]\rhodw_0-\jump[P_2^\dagger]\rhodw_0),\label{eqn:TPme}
\end{eqnarray}
where $\gamma'=\gamma_l+\gamma_r$, and we have used $\lind[B]\rho=\jump[B]\rho-\A[B]\rho$.  
The steady-state probability for the system to be in the singlet state is given by $\bra{S}\rho^{ss}\ket{S}=
\bra{S}\rho_0^{ss}+\rho_1^{ss}\ket{S}=1-(\coulb+\spenergydiff)^2/\DoubleOccupationEnergy^2$, is very close to unity.   Therefore, if the \DWS\ starts in a singlet state, its state does not change significantly during the measurement.  This further justifies the assertion that the measurement is nearly QND on the singlet subspace.

The poles of the master equation determine the measurement and relaxation rates.  There are poles at $0,-\gamma'\coulb^2/U^2,-\gamma'/2$ and $-\gamma'$.  The second of these corresponds to energy exchange processes generated by the operators $P_1$ and $P_5$ appearing in the Lindblad terms.  There is therefore a measurement-induced mixing time $\tmix=\frac{U^2}{\gamma'\coulb^2}$.  This mixing time is due to the fact that the singlet state is not quite an eigenstate of either $\hmeas$ nor $\hhub$.  
The mixing time is very long compared to the measurement time, $\sim1/\gamma'$, since $\coulb\ll U$.  The  mixing only induces relaxation of the \DWS, and so it has very little effect on the singlet state, which is already very close to the ground state.  We therefore conclude that this intrinsic mixing is negligible.


\subsection{SET average currents}\label{app:TPtraj}

To analyse the evolution of the measurement more formally, we unravel the unconditional master equation \eqn{eqn:2PMasterEquation}, and derive the conditional dynamics of quantum trajectories.  From the unravelling we can provide estimates for SET currents.  We assume the system may be described by the density matrix given in \eqn{eqn:R}, i.e.\ the SET island does not support coherent superpositions of $0$ and $1$ electrons.  We may therefore reduce the master equation given in \eqn{eqn:2PMasterEquation} to a pair of master equations for $\rhodw_0$ and $\rhodw_1$.  

The dynamics of the system decouple, depending on the  state of the \DWS.  In particular, the triplet states and $\ket{\Dminus}$ are eigenstates of the evolution operators, so we may consider the dynamics separately in each of the uncoupled, 1D subspaces $\{\ket{\tripup}\},\{\ket{\tripdown}\},\{\ket{\tripud}\},\{\ket{\Dminus}\}$.  In these subspaces the Hubbard Hamiltonian is proportional to the identity so the reduced master equations for each subspace are of the form
\begin{eqnarray}
\dot\rhodw_0&=&\gamma_r\jump[P]\rhodw_1-\gamma_l\A[P]\rhodw_0\\
\dot\rhodw_1&=&\gamma_l \jump[P]\rhodw_0-\gamma_r\A[P]\rhodw_1,
\end{eqnarray}
where $P$ is the projector onto the subspace, e.g.\ for the subspace $\{\ket{\Dminus}\}$, $P=\ket{\Dminus}\bra{\Dminus}$.

These reduced master equations depend on only a single jump operator $P$,  so the evolution between jumps may be written as a pure state, $\ket{\psi_\ns(t)}_c$, $\ns=0$ or 1, governed by the non-Hermitian effective Hamiltonian, $\heff_\ns$ according to the Schr\"odinger equation \cite{gar00}
\begin{equation}
\frac{d}{dt}\ket{\psi_\ns(t)}_c=-i \heff_\ns \ket{\psi_\ns(t)}_c.
\end{equation}
The non-Hermitian Hamiltonians for each subspace are
$
\heff_0=-i \frac{\gamma_l}{2}P \textrm{ and } \heff_1=-i \frac{\gamma_r}{2}P
$.
The solution is simply $\ket{\psi_0(t)}_c=e^{- \frac{\gamma_l}{2} t}\ket{\psi_0(0)}$ and $\ket{\psi_1(t)}_c=e^{- \frac{\gamma_r}{2} t}\ket{\psi_1(0)}$.  The jump rate is determined by the cumulative density function (CDF) for the waiting time between jumps $\prob_\ns(t_\jump<t)=1-{}_c\inner{\psi_\ns(t)}{\psi_\ns(t)}_c$ so $\prob_0(t_\jump<t)=1-e^{- \gamma_l t}$ and $\prob_1(t_\jump<t)=1-e^{- \gamma_r t}$.   Thus we have a cycle of electrons hopping onto the SET from the left lead at a rate $\gamma_l$ then hopping off to the right lead at a rate $\gamma_r$.  We therefore expect a current $i_T=e (\gamma_l^{-1}+\gamma_r^{-1})^{-1}$ to flow through the SET when the \DWS\ is in the triplet subspace, in agreement with above.

We now turn to the more complicated dynamics in the singlet subspace, $\{\ket{\singlet},\ket{\Dplus}\}$.  To simplify the analysis of this system, we will ignore the Lindblad terms depending on $P_1$ and $P_5$ in the master equation, \eqn{eqn:TPme}.  This is reasonable since these terms are $O(\coulb^2/U^2)$, which is small. For the resulting approximate form of \eqn{eqn:TPme}  the dynamics between jumps are governed by effective Hamiltonians for each SET island state, $\ns=0,1$, with a \emph{single} jump operator, 
$\heff_0=\hhub-i \frac{\gamma'}{2}P_2 P_2^\dagger$
 and 
 $\heff_1=\hhub+\hmeas-i \frac{\gamma'}{2}P_4^\dagger P_4$. 
Since there is only a single jump operator associated with  $\heff_\ns$,  we unravel the master equations as non-Hermitian Schr\"odinger equations for the pure, conditional, unnormalised, between-jump state-vectors, $\ket{\tilde\psi_\ns(t)}_c=\ueff_\ns(t)\ket{\psi_\ns(0)}$, where $\ueff_\ns(t)=e^{-i \heff_\ns t}$ \cite{gar00}.
During a jump at time $t_\jump$, the state of the system evolves discontinuously according to $\ket{\tilde\psi_1(t_\jump^+)}=P_2^\dagger\ket{\psi_0(t_\jump^-)}$ and $\ket{\tilde\psi_0(t_\jump^+)}=P_4\ket{\psi_1(t_\jump^-)}$.

To derive an upper bound on $i_\singlet$, we calculate the rate at which electrons hop on and off the SET, given that the \DWS\ begins in a singlet state.  The jump rate is determined by the CDF for the lifetime of the SET state with $\ns=0$ or 1 electrons, $\prob_\ns(t_\jump<t)=1-{}_c\inner{\psi_\ns(t)}{\psi_\ns(t)}_c$.  This quantity depends on the state of the \DWS\ immediately after the most recent jump, $\ket{\tilde\psi_{n_\isle}(t_\jump^+)}$, which is not deterministic due to the stochastic nature of the trajectory.  However, as discussed in section \ref{sec:singletmixing}, the steady state of the \DWS, $\rho^{ss}$, is very close to the singlet state since $\bra{
S}\rho^{ss}\ket{S}=1-(\coulb+\spenergydiff)^2/\DoubleOccupationEnergy^2$ \cite{gar00}.  Therefore, for the purposes of computing the CDF, to very good approximation, we can make the replacement $\ket{\tilde\psi_{n_\isle}(t_\jump^+)}\rightarrow\ket{\singlet}$.  The unnormalised, conditional state of the \DWS\ betweens jump is then 
\begin{eqnarray}
\ket{\tilde\psi_0(t)}_c&=&\ueff_0\ket{\singlet}\approx
e^{(i U-\gamma'/2)t} \ket{\singlet}+
O\left({\spenergydiff}/{\DoubleOccupationEnergy}\right),\nonumber\\
\ket{\tilde\psi_1(t)}_c&=&\ueff_1\ket{\singlet}\approx
e^{(i U- \frac{\gamma' {\spenergydiff }^2{\left( \coulb  + \spenergydiff  \right) }^2}{2\DoubleOccupationEnergy^4})t} \ket{\singlet}+
O\left(({\coulb+\spenergydiff})/{\DoubleOccupationEnergy}\right),\nonumber
\end{eqnarray}
where $t$ is measured from the previous jump.  It follows that the CDFs for the jump times are 
$
\prob_0(t_\jump<t)\approx1-e^{-\gamma' t}
$ and 
$
\prob_1(t_\jump<t)\approx1-e^{-\frac{ \spenergydiff^2{\left( \coulb  + \spenergydiff  \right) }^2}{\DoubleOccupationEnergy^4}\gamma' t}
$. 
These CDFs show that the lifetime of an empty SET is short, $\tau_0=1/\gamma'$, whilst the lifetime of an occupied SET is very long, $\tau_1=\frac{\DoubleOccupationEnergy^4}{ \spenergydiff^2{\left( \coulb  + \spenergydiff \right) }^2}\frac{1}{\gamma'}$.  Thus the cycle time for electrons to hop on and off the island is approximately $\tau_1$, and this provides an upper bound on the SET current when the \DWS\ starts in the singlet state, $i_\singlet<e/\tau_1$.  This is much less than $i_T$, in agreement with the heuristic prediction that $i_S=0$.

\begin{figure}
\subfigure[]{\includegraphics[width=5cm]{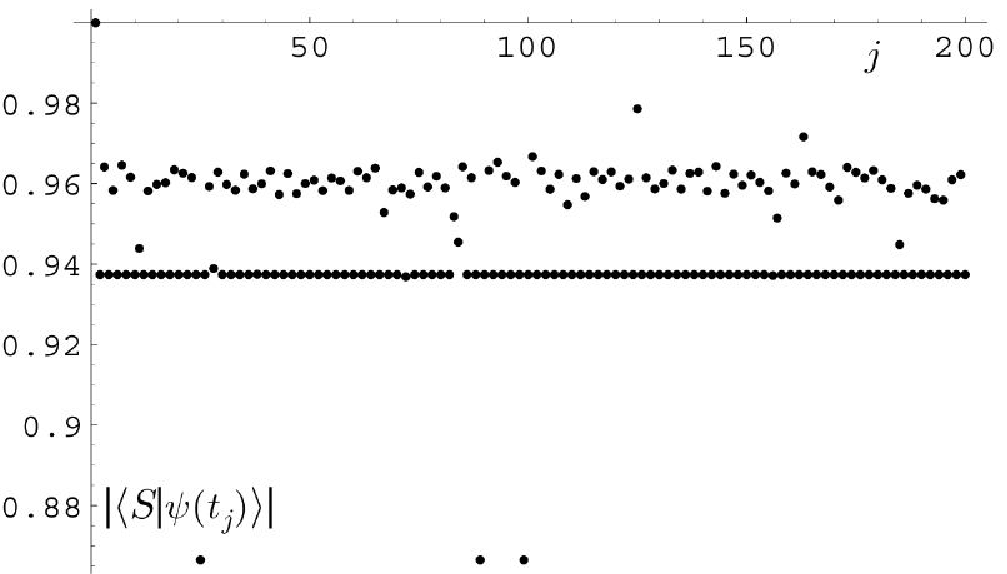}\label{fig:TwoParticleJumpTrace}}
\subfigure[]{\includegraphics[width=5cm]{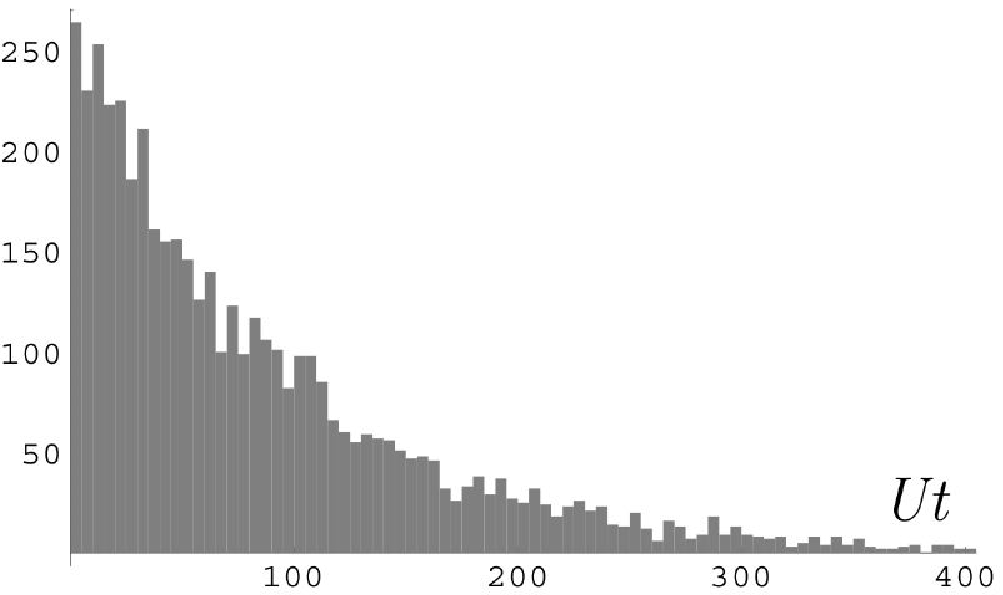}\label{fig:Histogram0}}
\subfigure[]{\includegraphics[width=5cm]{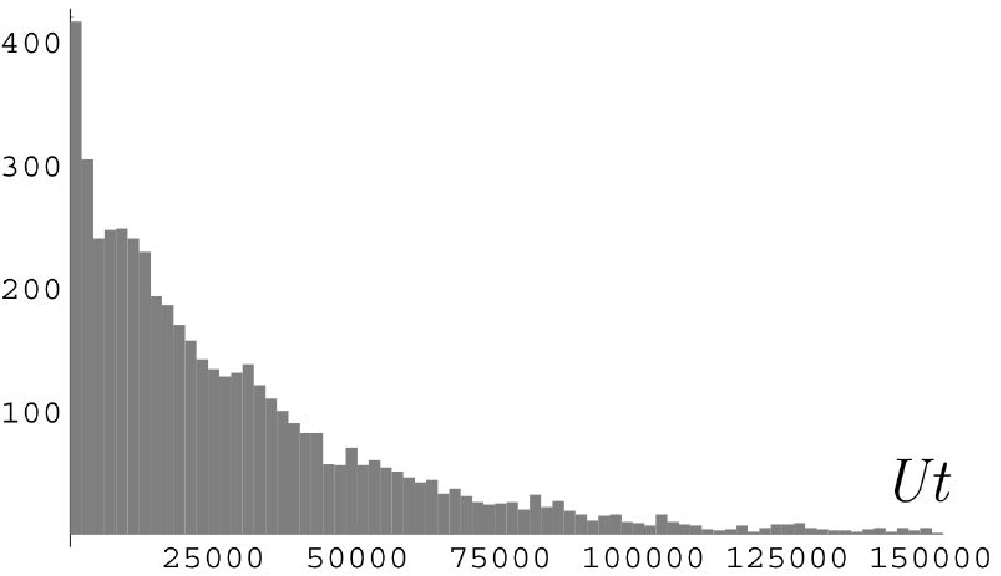}\label{fig:Histogram1}}
\caption{(a) Projection of the state of the \DWS\ onto the singlet state, $|\inner{\singlet}{\psi(t_j)}|$, immediately after jump $j$ for $\coulb=\spenergydiff/2=10\gamma'=\DoubleOccupationEnergy/10$.  
(b) Histogram of waiting times between jumps with $\ns=0$ and (c)  with $\ns=1$, for a simulation with $10^4$ jumps..  
}
\end{figure}

To confirm these approximate analytical predictions, we have performed numerical simulations of the conditional dynamics derived from the master equation \eqn{eqn:TPme}, which is not subject to any of the approximations made in this section.  In \fig{fig:TwoParticleJumpTrace} we show the projection $|\inner{\singlet}{\psi(t_j)}|$  immediately after jump $j$ for a sample of 200 jumps, assuming parameter values $\coulb=\spenergydiff/2=10\gamma'=\DoubleOccupationEnergy/10$. 
In this figure it is evident that the \DWS\ typically remains close to the singlet state after every jump, in agreement with the preceding analysis.  

Furthermore, we plot the distribution of waiting times between jumps for a simulation of 10000 sequential jumps, shown in \fig{fig:Histogram0} for an empty island ($\ns=0$) and in \fig{fig:Histogram1} for an occupied island ($\ns=1$).   For the parameters chosen, the figures show that the SET remains empty for a characteristic time of around $100/U$, whilst the typical occupation time is around $28000/U$.  These times are in agreement with the analytic estimates given above for which $\tau_0=100/U$ and $\tau_1=28000/U$.

This analysis establishes that since the \DWS\ remains close to the singlet state at all times, the measurement is close to an ideal QND measurement.  

\subsection{$B_k(t)$ for asymmetric \DWS}\label{app:assymBk}

We now consider the effect of asymmetry in the placement of the SET island, where $\coulc\neq0$.  
As in appendix \ref{app:noncomME} we express $B_k(t)$ in a Fourier decomposition restricted to the ordered basis $\{\ket{\singlet},\ket{\Dplus},\ket{\Dminus}\}$, which spans the \emph{singlet subspace}.  The Fourier components of this operator are crucial for deriving the master equation with which to analyse the system.  We find
\begin{eqnarray}
B_k(t)&=&e^{i\omega_k t}\Big(e^{i\DoubleOccupationEnergy t}\begin{bmatrix}
    0  & 0 & 0\\
    -\frac{ \coulb}{\DoubleOccupationEnergy}   &  0 & 0 \\
    0  & 0 & 0
\end{bmatrix}+
e^{i\frac{\coulb^2+2\coulb\spenergydiff}{\DoubleOccupationEnergy}t}\begin{bmatrix}
    1  & -\frac{ \coulb+\spenergydiff}{\DoubleOccupationEnergy} & 0 \\
    -\frac{ \spenergydiff}{\DoubleOccupationEnergy} &  0 & 0\\
    0 & 0& 0
\end{bmatrix}
+
e^{i2\coulc t}
\begin{bmatrix}
    0  & \frac{ \spenergydiff}{2\DoubleOccupationEnergy} & -\frac{ \spenergydiff}{2\DoubleOccupationEnergy}  \\
    \frac{ \coulb+\spenergydiff}{2\DoubleOccupationEnergy} &  \frac{1}{2} & - \frac{1}{2}  \\
    -\frac{ \coulb+\spenergydiff}{2\DoubleOccupationEnergy} &  -\frac{1}{2} &  \frac{1}{2}  
    \end{bmatrix}
\nonumber\\&&{}   +e^{-i2\coulc t}
\begin{bmatrix}
    0  & \frac{ \spenergydiff}{2\DoubleOccupationEnergy} & \frac{ \spenergydiff}{2\DoubleOccupationEnergy}  \\
    \frac{ \coulb+\spenergydiff}{2\DoubleOccupationEnergy} &  \frac{1}{2} &  \frac{1}{2}  \\
    \frac{ \coulb+\spenergydiff}{2\DoubleOccupationEnergy} &  \frac{1}{2} &  \frac{1}{2} 
    \end{bmatrix}
   +
 e^{-i\DoubleOccupationEnergy t}
\begin{bmatrix}
    0  & \frac{ \coulb}{\DoubleOccupationEnergy} & 0\\
    0   &  0 & 0\\
    0   &  0 & 0
\end{bmatrix}\Big).\label{eqn:FourierDecompBasym}
\end{eqnarray}
To derive this result, we have assumed that ${(\coulb+\spenergydiff)^2}/4\DoubleOccupationEnergy\ll{\coulc}\ll\DoubleOccupationEnergy$. In the opposite limit, ${(\coulb+\spenergydiff)^2}/4\DoubleOccupationEnergy\gg{\coulc}$, the effect of asymmetry is negligible. 
The energies $\omega_m$ appearing in the exponents above are shown schematically in \fig{fig:TwoParticleEnergyLevelAsym}.  As described in appendix \ref{app:lindop}, these energies correspond to the change in energy of an electron as it  tunnels between a lead and the SET, gaining or loosing energy as it interacts with the \DWS.   

\begin{figure}
\subfigure[]{\includegraphics{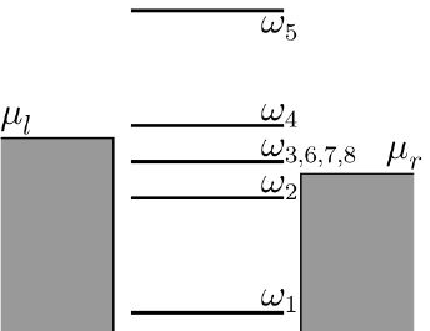}\label{fig:TwoParticleEnergyLevel}}\hfil
\subfigure[]{\includegraphics{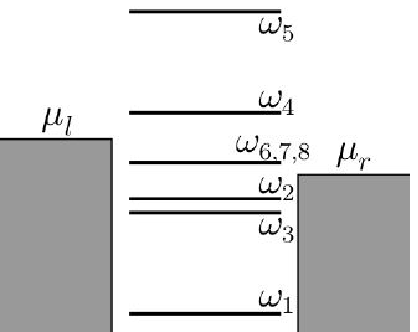}\label{fig:TwoParticleEnergyLevelAsym}}
\caption{(a) Transition energies, $\omega_m$, relative to lead Fermi levels, for a symmetrically placed SET island, where $\omega_1=-\omega_5=-\DoubleOccupationEnergy, \omega_2=-\omega_4=-(\coulb^2+2\coulb\spenergydiff)/U$ and $\omega_{3,6,7,8}=0$, and (b) for an asymmetrically placed SET island, where $\omega_1=-\omega_5=-\DoubleOccupationEnergy, \omega_2=-(\coulb^2+2\coulb\spenergydiff)/U, \omega_3=-\omega_4=-2\coulc$ and $\omega_{6,7,8}=0$.}
\end{figure}


\begin{thebibliography}{21}
\expandafter\ifx\csname natexlab\endcsname\relax\def\natexlab#1{#1}\fi
\expandafter\ifx\csname bibnamefont\endcsname\relax
  \def\bibnamefont#1{#1}\fi
\expandafter\ifx\csname bibfnamefont\endcsname\relax
  \def\bibfnamefont#1{#1}\fi
\expandafter\ifx\csname citenamefont\endcsname\relax
  \def\citenamefont#1{#1}\fi
\expandafter\ifx\csname url\endcsname\relax
  \def\url#1{\texttt{#1}}\fi
\expandafter\ifx\csname urlprefix\endcsname\relax\def\urlprefix{URL }\fi
\providecommand{\bibinfo}[2]{#2}
\providecommand{\eprint}[2][]{\url{#2}}

\bibitem[{\citenamefont{Aassime et~al.}(2001)\citenamefont{Aassime, Johansson,
  Wendin, Schoelkopf, and Delsing}}]{aas01}
\bibinfo{author}{\bibfnamefont{A.}~\bibnamefont{Aassime}},
  \bibinfo{author}{\bibfnamefont{G.}~\bibnamefont{Johansson}},
  \bibinfo{author}{\bibfnamefont{G.}~\bibnamefont{Wendin}},
  \bibinfo{author}{\bibfnamefont{R.~J.} \bibnamefont{Schoelkopf}},
  \bibnamefont{and} \bibinfo{author}{\bibfnamefont{P.}~\bibnamefont{Delsing}},
  \bibinfo{journal}{Phys. Rev. Lett.} \textbf{\bibinfo{volume}{86}},
  \bibinfo{pages}{3376} (\bibinfo{year}{2001}).

\bibitem[{\citenamefont{Makhlin et~al.}(2001)\citenamefont{Makhlin, Schon, and
  Shnirman}}]{mak01}
\bibinfo{author}{\bibfnamefont{Y.}~\bibnamefont{Makhlin}},
  \bibinfo{author}{\bibfnamefont{G.}~\bibnamefont{Schon}}, \bibnamefont{and}
  \bibinfo{author}{\bibfnamefont{A.}~\bibnamefont{Shnirman}},
  \bibinfo{journal}{Rev. Mod. Phys.} \textbf{\bibinfo{volume}{73}},
  \bibinfo{pages}{357} (\bibinfo{year}{2001}).

\bibitem[{\citenamefont{Buehler et~al.}(2003)\citenamefont{Buehler, Reilly,
  Starrett, Kenyon, Hamilton, Dzurak, and Clark}}]{bue03}
\bibinfo{author}{\bibfnamefont{T.~M.} \bibnamefont{Buehler}},
  \bibinfo{author}{\bibfnamefont{D.~J.} \bibnamefont{Reilly}},
  \bibinfo{author}{\bibfnamefont{R.~P.} \bibnamefont{Starrett}},
  \bibinfo{author}{\bibfnamefont{S.}~\bibnamefont{Kenyon}},
  \bibinfo{author}{\bibfnamefont{A.~R.} \bibnamefont{Hamilton}},
  \bibinfo{author}{\bibfnamefont{A.~S.} \bibnamefont{Dzurak}},
  \bibnamefont{and} \bibinfo{author}{\bibfnamefont{R.~G.} \bibnamefont{Clark}},
  \bibinfo{journal}{Microelectronic Engineering}
  \textbf{\bibinfo{volume}{67-8}}, \bibinfo{pages}{775} (\bibinfo{year}{2003}).

\bibitem[{\citenamefont{Stace et~al.}(2004)\citenamefont{Stace, Barnes, and
  Milburn}}]{sta04}
\bibinfo{author}{\bibfnamefont{T.~M.} \bibnamefont{Stace}},
  \bibinfo{author}{\bibfnamefont{C.~H.~W.} \bibnamefont{Barnes}},
  \bibnamefont{and} \bibinfo{author}{\bibfnamefont{G.~J.}
  \bibnamefont{Milburn}}, \bibinfo{journal}{cond-mat/0401442}
  (\bibinfo{year}{2004}).

\bibitem[{\citenamefont{Hayashi et~al.}(2003)\citenamefont{Hayashi, Fujisawa,
  Cheong, Jeong, and Hirayama}}]{hay03}
\bibinfo{author}{\bibfnamefont{T.}~\bibnamefont{Hayashi}},
  \bibinfo{author}{\bibfnamefont{T.}~\bibnamefont{Fujisawa}},
  \bibinfo{author}{\bibfnamefont{H.~D.} \bibnamefont{Cheong}},
  \bibinfo{author}{\bibfnamefont{Y.~H.} \bibnamefont{Jeong}}, \bibnamefont{and}
  \bibinfo{author}{\bibfnamefont{Y.}~\bibnamefont{Hirayama}},
  \bibinfo{journal}{Physical Review Letters} \textbf{\bibinfo{volume}{91}},
  \bibinfo{eid}{226804} (pages~\bibinfo{numpages}{4}) (\bibinfo{year}{2003}),
  \urlprefix\url{http://link.aps.org/abstract/PRL/v91/e226804}.

\bibitem[{\citenamefont{DiVincenzo et~al.}(2000)\citenamefont{DiVincenzo,
  Bacon, Kempe, Burkard, and Whaley}}]{div00}
\bibinfo{author}{\bibfnamefont{D.~P.} \bibnamefont{DiVincenzo}},
  \bibinfo{author}{\bibfnamefont{D.}~\bibnamefont{Bacon}},
  \bibinfo{author}{\bibfnamefont{J.}~\bibnamefont{Kempe}},
  \bibinfo{author}{\bibfnamefont{G.}~\bibnamefont{Burkard}}, \bibnamefont{and}
  \bibinfo{author}{\bibfnamefont{K.~B.} \bibnamefont{Whaley}},
  \bibinfo{journal}{Nature} \textbf{\bibinfo{volume}{408}},
  \bibinfo{pages}{339} (\bibinfo{year}{2000}).

\bibitem[{\citenamefont{Mahan}(1981)}]{mahan}
\bibinfo{author}{\bibfnamefont{G.~D.} \bibnamefont{Mahan}},
  \emph{\bibinfo{title}{Many-Particle Physics}} (\bibinfo{publisher}{Plenum
  Press, New York}, \bibinfo{year}{1981}).

\bibitem[{\citenamefont{Kane}(1998)}]{kan98}
\bibinfo{author}{\bibfnamefont{B.~E.} \bibnamefont{Kane}},
  \bibinfo{journal}{Nature} \textbf{\bibinfo{volume}{393}},
  \bibinfo{pages}{133} (\bibinfo{year}{1998}).

\bibitem[{NaS(2003)}]{NaSi}
\emph{\bibinfo{title}{Silicon Quantum Information Processing, Project
  \#DAAD19-01-1-0552}} (\bibinfo{organization}{ARO Quantum Computing Program
  Review}, \bibinfo{year}{2003}).

\bibitem[{\citenamefont{Burkard et~al.}(1999)\citenamefont{Burkard, Loss, and
  DiVincenzo}}]{bur99}
\bibinfo{author}{\bibfnamefont{G.}~\bibnamefont{Burkard}},
  \bibinfo{author}{\bibfnamefont{D.}~\bibnamefont{Loss}}, \bibnamefont{and}
  \bibinfo{author}{\bibfnamefont{D.~P.} \bibnamefont{DiVincenzo}},
  \bibinfo{journal}{Phys. Rev. B} \textbf{\bibinfo{volume}{59}},
  \bibinfo{pages}{2070} (\bibinfo{year}{1999}).

\bibitem[{\citenamefont{Schofield et~al.}(2003)\citenamefont{Schofield, Curson,
  Simmons, Ruess, Hallam, Oberbeck, and Clark}}]{sch03}
\bibinfo{author}{\bibfnamefont{S.~R.} \bibnamefont{Schofield}},
  \bibinfo{author}{\bibfnamefont{N.~J.} \bibnamefont{Curson}},
  \bibinfo{author}{\bibfnamefont{M.~Y.} \bibnamefont{Simmons}},
  \bibinfo{author}{\bibfnamefont{F.~J.} \bibnamefont{Ruess}},
  \bibinfo{author}{\bibfnamefont{T.}~\bibnamefont{Hallam}},
  \bibinfo{author}{\bibfnamefont{L.}~\bibnamefont{Oberbeck}}, \bibnamefont{and}
  \bibinfo{author}{\bibfnamefont{R.~G.} \bibnamefont{Clark}}
  (\bibinfo{year}{2003}), \eprint{arXiv:cond-mat/0307599}.

\bibitem[{\citenamefont{Barrett and Stace}()}]{SeansPaper}
\bibinfo{author}{\bibfnamefont{S.~D.} \bibnamefont{Barrett}} \bibnamefont{and}
  \bibinfo{author}{\bibfnamefont{T.~M.} \bibnamefont{Stace}}, \bibinfo{note}{in
  preparation}.

\bibitem[{\citenamefont{Stace and Barrett}(2004)}]{stace:136802}
\bibinfo{author}{\bibfnamefont{T.~M.} \bibnamefont{Stace}} \bibnamefont{and}
  \bibinfo{author}{\bibfnamefont{S.~D.} \bibnamefont{Barrett}},
  \bibinfo{journal}{Phys. Rev. Lett.} \textbf{\bibinfo{volume}{92}},
  \bibinfo{eid}{136802} (pages~\bibinfo{numpages}{4}) (\bibinfo{year}{2004}),
  \urlprefix\url{http://link.aps.org/abstract/PRL/v92/e136802}.

\bibitem[{\citenamefont{Fl{\"u}gge}(1999)}]{flu99}
\bibinfo{author}{\bibfnamefont{S.}~\bibnamefont{Fl{\"u}gge}},
  \emph{\bibinfo{title}{Practical Quantum Mechanics}}
  (\bibinfo{publisher}{Springer, Berlin}, \bibinfo{year}{1999}).

\bibitem[{\citenamefont{Vrijen et~al.}(2000)\citenamefont{Vrijen, Yablonovitch,
  Wang, Jiang, Balandin, Roychowdhury, Mor, and DiVincenzo}}]{vri00a}
\bibinfo{author}{\bibfnamefont{R.}~\bibnamefont{Vrijen}},
  \bibinfo{author}{\bibfnamefont{E.}~\bibnamefont{Yablonovitch}},
  \bibinfo{author}{\bibfnamefont{K.}~\bibnamefont{Wang}},
  \bibinfo{author}{\bibfnamefont{H.~W.} \bibnamefont{Jiang}},
  \bibinfo{author}{\bibfnamefont{A.}~\bibnamefont{Balandin}},
  \bibinfo{author}{\bibfnamefont{V.}~\bibnamefont{Roychowdhury}},
  \bibinfo{author}{\bibfnamefont{T.}~\bibnamefont{Mor}}, \bibnamefont{and}
  \bibinfo{author}{\bibfnamefont{D.}~\bibnamefont{DiVincenzo}},
  \bibinfo{journal}{Phys. Rev. A} \textbf{\bibinfo{volume}{62}},
  \bibinfo{pages}{012306} (\bibinfo{year}{2000}).

\bibitem[{\citenamefont{Barrett and Milburn}(2003)}]{bar03}
\bibinfo{author}{\bibfnamefont{S.~D.} \bibnamefont{Barrett}} \bibnamefont{and}
  \bibinfo{author}{\bibfnamefont{G.~J.} \bibnamefont{Milburn}}
  (\bibinfo{year}{2003}), \eprint{arXiv:cond-mat/0302238}.

\bibitem[{\citenamefont{Furusaki and Matveev}(1995)}]{fur95}
\bibinfo{author}{\bibfnamefont{A.}~\bibnamefont{Furusaki}} \bibnamefont{and}
  \bibinfo{author}{\bibfnamefont{K.~A.} \bibnamefont{Matveev}},
  \bibinfo{journal}{Phys. Rev. B} \textbf{\bibinfo{volume}{52}},
  \bibinfo{pages}{16676} (\bibinfo{year}{1995}).

\bibitem[{\citenamefont{Datta}(1995)}]{dat95}
\bibinfo{author}{\bibfnamefont{S.}~\bibnamefont{Datta}},
  \emph{\bibinfo{title}{Electronic Transport in Mesoscopic Systems}}
  (\bibinfo{publisher}{Cambridge University Press},
  \bibinfo{address}{Cambridge}, \bibinfo{year}{1995}).

\bibitem[{\citenamefont{Wiseman et~al.}(2001)\citenamefont{Wiseman, Utami, Sun,
  Milburn, Kane, Dzurak, and Clark}}]{wis01}
\bibinfo{author}{\bibfnamefont{H.~M.} \bibnamefont{Wiseman}},
  \bibinfo{author}{\bibfnamefont{D.~W.} \bibnamefont{Utami}},
  \bibinfo{author}{\bibfnamefont{H.~B.} \bibnamefont{Sun}},
  \bibinfo{author}{\bibfnamefont{G.~J.} \bibnamefont{Milburn}},
  \bibinfo{author}{\bibfnamefont{B.~E.} \bibnamefont{Kane}},
  \bibinfo{author}{\bibfnamefont{A.}~\bibnamefont{Dzurak}}, \bibnamefont{and}
  \bibinfo{author}{\bibfnamefont{R.~G.} \bibnamefont{Clark}},
  \bibinfo{journal}{Phys. Rev. B} \textbf{\bibinfo{volume}{63}},
  \bibinfo{pages}{235308} (\bibinfo{year}{2001}).

\bibitem[{\citenamefont{Sridhar and Jagannathan}(2002)}]{sri02}
\bibinfo{author}{\bibfnamefont{R.}~\bibnamefont{Sridhar}} \bibnamefont{and}
  \bibinfo{author}{\bibfnamefont{R.}~\bibnamefont{Jagannathan}}
  (\bibinfo{year}{2002}), \eprint{math-ph/0212068}.

\bibitem[{\citenamefont{Gardiner and Zoller}(2000)}]{gar00}
\bibinfo{author}{\bibfnamefont{C.~W.} \bibnamefont{Gardiner}} \bibnamefont{and}
  \bibinfo{author}{\bibfnamefont{P.}~\bibnamefont{Zoller}},
  \emph{\bibinfo{title}{Quantum Noise}} (\bibinfo{publisher}{Springer},
  \bibinfo{year}{2000}).

\end{thebibliography}
\renewcommand{\url}[1]{}
\newcommand{\urlprefix}[1]{#1}

\end{document}